\documentclass{aa}
\usepackage{txfonts}
\usepackage{graphicx}
\usepackage{rotating}
\usepackage{lscape}

\begin{document}

\title{CRIRES-POP\thanks{Based on observations collected at the European 
Organisation for Astronomical Research in the Southern Hemisphere, Chile 
(084.D-0912, 085.D-0161, 086.D-0066, and 087.D-0195. The spectra presented
in Figs.~3 to 15 are available at the CDS via anonymous ftp to 
cdsarc.u-strasbg.fr (130.79.128.5) or via http://cdsweb.u-strasbg.fr/cgi-bin/qcat?J/A+A/}}
\subtitle{A library of high resolution spectra in the near-infrared}

\author{
T. Lebzelter\inst{1} 
\and A. Seifahrt\inst{2,9}
\and S. Uttenthaler\inst{1}
\and S. Ramsay\inst{3}
\and H. Hartman\inst{4}
\and M.-F. Nieva\inst{5,6}
\and N. Przybilla\inst{6}
\and A. Smette\inst{3}
\and G.M. Wahlgren\inst{7}
\and B. Wolff\inst{3}
\and G.A.J. Hussain\inst{3}
\and H.U. K\"aufl\inst{3}
\and U. Seemann\inst{3,8}
}

\institute{
Department of Astronomy, University of Vienna, T{\"u}rkenschanzstra{\ss}e 17, A-1180 Vienna, Austria
\and Physics Department, University of California, One Shields Avenue, Davis, CA 95616, USA
\and ESO, Karl-Schwarzschild-Str. 2, D-85748 Garching b. M\"unchen, Germany
\and Lund Observatory, Lund University, Box 43, SE-22100 Lund, Sweden
\and Max-Planck-Institute for Astrophysics, Karl-Schwarzschild-Str. 1, D-85741 Garching b. M\"unchen, Germany
\and Dr. Karl Remeis-Observatory \& ECAP, University Erlangen-Nuremberg,
Sternwartstr. 7, D-96049 Bamberg, Germany
\and Dept. of Physics, The Catholic University of America, 620 Michigan Ave NE, Washington DC, USA 20064
\and Institut f\"ur Astrophysik, Georg-August-Universit\"at G\"ottingen, Friedrich-Hund-Platz 1, 37077 G\"ottingen, Germany
\and Department of Astronomy and Astrophysics, University of Chicago, 5640 S. Ellis Ave, Chicago, IL 60637, USA
}

\date{Received / Accepted}

\abstract{New instrumental capabilities and the wealth of astrophysical information extractable from
the near-infrared wavelength region have led to a growing interest in the field of high resolution
spectroscopy at 1--5\,$\mu$m.}
{We aim to provide a library of observed high-resolution and high signal-to-noise-ratio near-infrared spectra of stars of various
types throughout the Hertzsprung-Russell diagram. This is needed for the exploration of spectral features in
this wavelength range and for comparison of reference targets with observations and models.}
{High quality spectra were obtained using the CRIRES near-infrared spectrograph
at ESO's VLT covering the
range from 0.97\,$\mu$m to 5.3\,$\mu$m at high spectral resolution. Accurate wavelength calibration and correction for
of telluric lines were performed by fitting synthetic transmission
spectra for the Earth's atmosphere to each spectrum individually.}
{We describe the observational strategy and the current status and content of the library which includes
13 objects. The first examples of finally reduced spectra are presented. This publication will serve as a reference paper
to introduce the library to the community and explore the extensive amount of material.}
{}


\maketitle

\section{Introduction} 
After a long-term reliance on the visual spectral region, it has been clearly demonstrated in the past two decades that the near- and mid-infrared spectral ranges will play a leading role in many prime areas of astronomical research 
for the coming decades.
This part of the spectrum enables observation of a universe of "cool" phenomena, including disks, planets, or the extended atmospheres of evolved 
stars. Electronic transitions of atoms, and in the case of molecules rotation-vibration transitions, produce 
lines in the near-infrared (NIR), among them indicators for obtaining isotopic ratios and resonance lines of {\it s}-process and rare earth elements. For the coolest stars, the optical spectrum is dominated by molecular lines, limiting 
the extent to which chemical compositions can be determined owing to blending. 
At infrared (IR) wavelengths there are spectral windows that are relatively free of 
molecular absorption, where lines from heavy elements are located. Furthermore, the reduced effects
of extinction in the NIR compared to the visual range allow us to access more distant stars. 

To fully understand the information contained in the NIR spectral range, high-resolution spectroscopy is mandatory. Considerable 
progress in instrument development has been achieved over the past twenty years, resulting in the current availability of efficient 
spectrographs for the NIR. 
The lack of high quality (high spectral resolution and high signal-to-noise ratio -- 
S/N) stellar spectra in 
the IR has meant that spectroscopists have concentrated their resources on the strongest features observed at 
lower spectral resolution in the IR or in other wavelength regions. As a result, line data for 
many elements, but especially the post-iron group elements, 
are severely lacking and those that exist are predominantly the result of theoretical calculations.  
The atlases of the solar spectrum (Wallace et al. \cite{Wallace96}, Delbouille et al. \cite{DRN73}) 
and the K giant Arcturus (Hinkle et al. 
\cite{arcturus}) illustrated the value of a stellar reference spectrum at high spectral resolution and high S/N. 

At high resolving power ($R = 100\,000$), an inventory of complete NIR stellar spectra across the Hertzsprung-Russell  diagram (HRD)  is presently unavailable. Spectra of both
high spectral resolution and high S/N are key demands for a list of
primary science cases in addition to the above-mentioned exploration of weak atomic lines of hardly studied elements.
In the case of molecular lines, both prerequisites are needed to study isotopic ratios, which open a window to
the nuclear fusion processes in the stellar intererior. Another topic that can be investigated with the help 
of such data are small velocity shifts between high and low 
excitation lines or asymmetries in the line profiles 
that hint towards velocity fields in a star's atmosphere. Weak emission features that need to be clearly
distinguished from absorption components are promising tools to study circumstellar material. Even for observations
at lower resolution, the availability of a high resolution reference library is required to interpret the
results.

We therefore started to
build such a library of complete (full wavelength coverage from 1  to 5\,$\mu$m), high S/N NIR spectra of  representatives of various parts of the HRD using
the  spectrograph CRIRES (CRyogenic high-resolution InfraRed Echelle Spectrograph; K\"aufl et al. \cite{crires}) at ESO's VLT. With our chosen setup, it will be possible to study lines  
fainter than 1\% of the continuum in strength, and to detect line asymmetries at this level.
The availability of this library provides an unprecedented database for the testing of
stellar model atmopheres, the improvement of line lists of various molecules and atoms, the identification of weak lines, and the proper correction of telluric absorption
features.  At this point, an almost complete wavelength coverage has been secured for 13 stars. The total library will consist of about 25 objects. Further data will become available
continuously via the dedicated webpage of the project\footnote{\it  http://www.univie.ac.at/crirespop}. The intention of this paper is to introduce the library to the astronomical
community, discuss aspects of the data reduction, present examples, and provide a reference for future exploitation of the library spectra.

\section{Observing strategy and instrument settings} 
CRIRES is a NIR high-resolution echelle spectrograph operating
at from 0.96 to 5.3 $\mu$m. An adaptive optics (AO) system using a curvature wave 
front
sensor can be utilized during observations. The instrument works in a near-Littrow configuration. 
Variable entrance slit widths
from 0.05 to 3.0 arcsec are offered,
although in practice only the 0.2 to 0.4 arcsec slits are typically used.
Active guiding on the slit image in a NIR filter ($J$, $H$, or $K$, matching as close
as possible the wavelength setting of the spectrograph) is part of the standard observing mode.
CRIRES covers a comparably small wavelength range of a few 
tens of nm per observational setting (depending on the  central
wavelength). To observe the entire range from below 1 to above 5\,$\mu$m thus requires about 200 settings and a considerable amount of observing time. 
A complete scan of a star
with $K=1$\,mag reaching a S/N of at least 200 throughout the entire spectral range takes almost nine hours and is strongly dominated by 
observational overheads (close to 80\,\%).  To achieve our goal of a library
covering all main spectral types and luminosity classes, we  limited the list of possible targets to rather bright sources ($K<4\fm5$), and designed the project as a filler programme
at the Very Large Telescope (VLT) capable of using observing time with non-photometric seeing conditions.

The selection of the targets was driven by the following criteria: they are (1) representative of a specific spectral type 
and luminosity class or a specific abundance pattern; (2) of sufficient brightness in the CRIRES range; 
(3) of very low 
rotational velocity (typically less than 10\,km\,s$^{-1}$); (4) found to have no large amplitude variability  
(less than 1 mag in $V$); (5) part of the UVES-POP\footnote{UVES-POP is an electronic library of 
high-resolution optical spectra obtained with UVES, the Ultraviolet-Visual Echelle Spectrograph at the VLT in the
framework of a Paranal Observatory Project.} sample (Bagnulo et al. \cite{bagnulo03}) allowing
to achieve complete 
wavelength coverage from 300\,nm to 5\,$\mu$m . The latter criterion could not be fulfilled in all cases owing to the 
lack of stars suitable for observations in the NIR range. Furthermore, no suitable target exists to 
represent spectral type O, or Wolf-Rayet stars at NIR wavelengths. 
We included targets with non-standard chemistry, such as a carbon star and a chemically
peculiar Ap star. The list of stars observed to date can be found in Table \ref{t:sample}, and a brief description of each star is given in Sect.\,\ref{library}. Observations continue to be performed, 
and a complete
list of the stars in the library will be presented elsewhere at a later date.

All wavelength settings between 950\,nm and 5300\,nm 
offered by CRIRES were observed, except for those located in-between the atmospheric windows,
namely 1350--1450 nm, 1800--1984 nm, 2510--3425 nm, and 4190--4620 nm.
This provided a list of approximately 200 settings per target. The wavelength ranges of 
the settings overlapped with each other to close the gaps between the detector chips. This had the positive side effect 
of increasing the S/N of the final data and allowing us to check for instrumental 
problems. For a few hot stars, we had not yet observed the $M$-band, because we decided to take a closer look at the 
$M$-band spectrum of one hot star first and identify the most promising wavelength settings for observations. This would allow us to optimise the use of our observing time.

For CRIRES-POP, we chose a slit width of 0.2 arcsec, which gives
a measured resolving power at 2172 nm of $R = 96\,000$ (Smoker \cite{smoker11}). 
The AO system was not used because the actual spectral resolution
is determined by the entrance slit and does not depend on the AO correction.

A pre-dispersion prism and an intermediate slit are used within CRIRES
to limit the wavelength range to a single order
before light enters the high-resolution spectrograph, which consists of a
31.6 lines/mm, 63.5 degrees blaze echelle grating. The science detector is a 
4x1 mosaic of Aladdin III InSb arrays. In total, 4096x512 pixels are read-out;
the gap between two successive arrays is equivalent to 280 pixels.
The detector has several defects, particularly a diagonal
scratch on the right part of array 2, a vertical line of bad pixels
near the centre of array 4, and a larger number of bad pixels at the
very left of array 4. Almost all of these cosmetic defects will be corrected in
the final data set by using spectra obtained at different nod positions and 
overlapping wavelength settings.

There is also an odd-even like pattern present that
is aligned with the read-out direction. This is in the dispersion direction
for arrays 1 and 4 and in cross-dispersion direction for arrays 2 and 3. The
effect can be almost completely removed (with residuals in the extracted flux
of $\ll$ 1\%) by flat-fielding and applying
a correction for detector non-linearity (see Smoker \cite{smoker11}, Jung \cite{Jung10},
Uttenthaler \cite{Uttenthaler07}).

At shorter wavelengths (typically below 1.2 $\mu$m), an optical ghost is occasionally
present, which is caused by a reflection from the science detector back to
the grating. It often does not cancel out completely via nodding and is
visible as an artificial absorption feature in extracted spectra.

Integration times were chosen in accordance with the CRIRES exposure time calculator offered by ESO 
to achieve a S/N of at least 200 in average seeing conditions. 
Observations were always done in an ABBA nodding pattern with typically a 15" nod throw
to account for the background emission. Below 2.5\,$\mu$m,
total integration times per setting
were between 10\,s for the brightest stars and 120\,s for the faintest. In the $L$-band,
total integration times were between 60\,s and 360\,s. For the weaker stars, this was split into several integrations
at each nodding position and up to six nodding cycles. Finally, for the $M$-band, integration times of up to 1200\,s
were used, again split into several integrations and nodding cycles because of 
the high thermal background 
contribution. Between 1 and 23 wavelength settings were combined into observing blocks of 20 to 60 minutes length. We used the no-AO mode. 
Calibration data were taken during day time after each observing night in which our
program was executed.

Since CRIRES-POP was designed as a filler programme, the various wavelength ranges of a given star 
were observed over up to 
1.5 years. A requirement for the selection was, thus, that a target was of 
constant brightness, or only a mildly variable star. 
No airmass limit was set, although we note that 
only a small fraction of observations were obtained at an airmass
above 2.

\section{Data reduction}
\subsection{First step: standard CRIRES pipeline} \label{extract}
Our CRIRES-POP observations were reduced with recipes available from the ESO CRIRES data reduction pipeline 2.0.0
(Jung \cite{Jung10}). Raw science frames were flat-fielded, and corrected for both non-linear detector response and bad pixels. 
Frames from different nodding positions were shifted and combined. Finally,
the spectrum was extracted and wavelength calibrated. The 
wavelength calibration was determined from either daytime ThAr 
arc lamp exposures for settings below 2.4~$\mu$m or sky emission lines in the science exposures (above 
2.4~$\mu$m). 
We found a few instrumental artefacts that in our data and that could not be removed 
during the reduction process. Neighbouring
echelle orders partially overlap at shorter wavelengths, which means that part of the spectrum for a 
given order can be contaminated by adjacent orders. This is essentially true below 1.3~$\mu$m, where data from 
detectors 
1 and 4 cannot be used.  
When observing spectral regions with fully saturated telluric features,
the actually observed line profiles do typically show 1-3\,\% of residual
flux in the absorption cores. This is due to in-dispersion straylight
produced by the CRIRES grating. The effect can be modelled quantitatively
based on line-profiles obtained with deep integrations of neon or
krypton arc calibration lamps when the entrance slit is illuminated
uniformly. When observing a star, i.e.~a point-like source, the
spatial distribution of the scattered light flux in the cross-dispersion
direction along the slit then differs from the normal spatial profile of
the stellar continuum. Thus, the optimum extraction algorithm employed
cannot easily extract precise fluxes from the center of absorption
lines. Moreover, owing to the field-dependent tilt between straylight
direction and the stellar continuum, there is no easy way to model the
effect. Hence, the effect cannot be corrected for yet, regardless of whether it is caused by telluric
absorption or absorption lines in the stellar atmosphere.

Wavelength calibration is a difficult task for CRIRES because of the rather small spectral ranges that are 
measured 
on each detector. Wavelength calibration was performed separately for each detector and required the availability of a 
sufficient number of calibration lines in the respective spectral region. This was not always the case, especially when 
a ThAr exposure was used for calibration. Then a default linear solution was applied. The situation, however, might have been 
improved by using telluric lines (see below).

Small differences in the wavelength calibration of the two nodded images resulted 
in a slight degradation of the spectral resolution of the combined image
that  depended on the nodding amplitude and specific wavelength setting. We verified on actual  data that this broadening was smaller than two pixels
(i.e. 5.10$^{-3}$ nm/pixel at 1~$\mu$m or 25.10$^{-3}$ nm/pixel at 5~$\mu$m) for the  observing strategy used for 
CRIRES-POP and that the final spectral resolution was on average
degraded by about 10--20\% (R$=$80\,000 to 90\,000). Future releases of the CRIRES-POP data will use a new CRIRES pipeline recently released 
by ESO (version 2.1.3), that is capable of
reducing spectra from A and B nodding positions separately. 
In the future we will make use of this feature to reduce spectra where 
the degradation to the resolution
introduced by the slight curvature limits the final data quality.

\subsection{Second step: telluric correction using a model of Earth's atmosphere} \label{telluric}
A general problem 
in almost all NIR spectra is telluric absorption lines. Traditionally, they are divided out of a 
science spectrum using a hot (i.e. feature-poor) "standard" star spectrum. In the case of CRIRES-POP, 
observations of telluric standards would have been impractical for two reasons. 
First, it would have led to an enormous amount of calibration measurements, taking into account the large number of 
wavelength settings involved and the need to observe the target and the standard star close in time and airmass. 
Second, telluric standard stars contain a number of spectral features, some of them unidentified, such that they do not provide ideal
'featureless' spectra and would compromise the spectra of our science targets. This is also a reason why we 
included early-type stars in our program, i.e. to explore the intrinsic features of these stars in the 
IR. Thus, we decided against observing telluric standard stars but chose a different approach, to model the 
telluric spectrum at the time of observation. 

Model spectra of the telluric transmission can be calculated with freely available radiative transfer codes (e.g. the 
Line-by-Line Radiative Transfer Model [LBLRTM] or the Reference Forward Model [RFM]), using a database of molecular 
transition parameters (e.g., HITRAN, the HIgh-Resolution TRANsmission molecular absorption database)
and atmospheric profiles (e.g. 
GDAS\footnote{Global Data Assimilation System \texttt{http://ready.arl.noaa.gov/READYamet.php}}). The modelling of CRIRES spectra, as described in Seifahrt 
et al. (\cite{Sei10}) 
and Smette et al. (\cite{smette10}), involves the modelling of the instrumental profile, the scaling of the column 
densities of the trace gases in the Earth's atmosphere, as well as a wavelength solution for each CRIRES chip as a 
second or third order polynomial. Hence, this approach has the additional advantage of providing us with a precise 
wavelength calibration that is often difficult to obtain with standard calibration techniques. 

For early-type stars with a clear continuum and few spectral features, fitting of telluric absorption models is semi-automatic. Late-type stars with an
abundance of stellar lines require a more restricted type of 
modelling and manual intervention caused by the strong blending of telluric and stellar features. In
extreme cases, the instrumental profile and abundance scaling factors have to be fixed based on observations of early-type stars in the same night and
only a solution of the wavelength vector and the most variable  features (i.e. water vapor) can be obtained for the target. Overlapping wavelength regions on different
chips obtained in  neighbouring settings provide an important sanity check to the modelling in these cases.

The current  standard of line data of relevant molecules
in the Earth's atmosphere is set by the HITRAN consortium (Rothman et al. \cite{Roth08}). We 
were able to confront 
the model spectra with our high-quality CRIRES spectra
(of hot stars, e.g. $\tau$\,Sco, e Vel), to test the quality of the available molecular data. Similar 
tests of molecular data had already been done by  some of us for a
few selected CRIRES wavelength settings (Seifahrt et al. \cite{Sei10};  Smette et al. \cite{smette10}). With the current  CRIRES-POP data base, however, it
becomes feasible to test molecular data for almost the entire wavelength range  between 0.97
$\mu$m and 5.3 $\mu$m (a few small ranges were omitted because they were affected by heavy
telluric absorption). We are in contact  with the HITRAN consortium to work out a procedure for a potential improvement of
the line data. In a first step, the poorly-fit
lines will be catalogued. If there is a line that deviates from its  observed position or strength, it can be flagged as suspicious. The HITRAN list has quality
indicators for each line  in the database, that can be used to evaluate whether the observed deviation is expected. If the line is  consistently off by
more than the expected uncertainty listed in HITRAN in several observed spectra, this line  deserves further scrutiny. The wealth of information contained in the
CRIRES-POP database turns this effort to improve molecular line data 
into a considerable project of its own, which will have a large legacy value.
Meanwhile, a moderate amount of back-fitting of line parameters in terms of position and strength can improve the  telluric fitting for our project.
For the final data set, we aim to provide telluric corrections for the
entire wavelength range covered by our observations where telluric
lines are detected at the 2 $\sigma$ level. We expect the estimated
performance of the correction to be at or above the level shown in
Seifahrt et al. (\cite{Sei10}), especially when the discussed back-fitting of line
parameters is applied.
\subsection{Combining individual spectral settings}
Once all individual one-dimensional (1D) 
spectra from a single star had been extracted (Sec. \ref{extract}) 
and cleaned of telluric absorption features (Sec. \ref{telluric}), 
we combined those spectra into a single data product. At 
this step, we used the spectral overlap of the individual settings to improve the final SNR, 
and to exclude regions
that show contamination from neighbouring orders, optical ghosts, or other data defects (e.g. bad pixels). For 
spectral regions that include telluric absoption lines, we used the superior wavelength calibration delivered 
by the telluric modelling and combine all spectra after mapping them onto a common wavelength vector.
At this stage, we also took into account any velocity shifts resulting from obtaining the various parts
of the spectra at different epochs. The combination of several adjacent settings into one hour observing blocks
ensured that in most cases only small corrections were needed. 
For the final library the spectra will be shifted in wavelength to
zero heliocentric velocity.

\section{Library content} \label{library}
In the following sections, we summarised the information from 
the literature about the global parameters and any determined atmospheric
parameters for these stars.\footnote{The 
latest version of the CRIRES data for these stars is available via {\it 
http://www.univie.ac.at/crirespop}.} The final release of the data will consist of the raw, pipeline-reduced, and 
telluric-absorption-corrected spectra for the
complete observed sample.

\begin{table*}
\caption{Cross references and spectral types of the stars currently
in the CRIRES-POP library, sorted by spectral type. The last column gives the year
of the observations.}
\label{t:sample}
\centering
\begin{tabular}{llcrrccl}
\hline\hline
HD number & other names & spectral type & RA (2000) & DE (2000) & $K$ & UVES-POP?\tablefootmark{a} & observing\\
          &             &               &           &           & (mag) &                          &  periods \\
\hline
149438 & \object{$\tau$\,Sco}, HR\,6165 & B0.2 V & 16:35:52 & $-$28:12:58 & 3.60 & n & 2007\tablefootmark{b} / 2009$-$2011\\
47105 & \object{$\gamma$ Gem}, HR\,2421 & A0 IV & 06:37:42 & +16:23:57 & 1.92 & n & 2010\\
118022\tablefootmark{c} & $o$ Vir, HR\,5105\,CW Vir\,78 Vir & A1p & 13:34:08 & +03:39:32 & 4.88 &  y & 2010$-$2011\\
73634 & \object{e Vel}, HR\,3426 & A6 II & 08:37:39 & $-$42:59:21 & 3.6 & y & 2009$-$2010\\
20010 & \object{$\alpha$ For}, HR\,963, LHS 1515 & F8V & 03:12:05 & $-$28:59:15 & 2.32 & y & 2009$-$2010\\
109379 & \object{$\beta$ Crv}, HR\,4786 & G5 II & 12:34:23 & $-$23:23:48 & 0.8 & y & 2010$-$2011\\
83240 & \object{10\,Leo}, HR\,3827 & K1 III & 09:37:13 & +06:50:09 & 2.66 & y & 2009$-$2010\\
225212 & \object{3\,Cet}, HR\,9103 & K3 I & 00:04:30 & $-$10:30:34 & 1.4 & y & 2010$-$2011\\
49331 & \object{HR\,2508} & M1 I & 06:47:37 & $-$08:59:55 & 0.6 & y & 2010$-$2011\\
224935 & HR\,9089, \object{YY Psc} & M3 III & 00:01:58 & $-$06:00:51 & -0.5 & y & 2009$-$2010\\
 & Barnard's star, \object{GJ 699}, V2500 Oph & M4 V & 17:57:48 & +04:41:36 & 4.52 &  n & 2010$-$2011\\
61913 & HR\,2967, \object{NZ Gem} & S & 07:42:03 & +14:12:31 & 0.56 & y & 2010$-$2011\\
134453 & HR\,5644, \object{X TrA} & C & 15:14:19 & $-$70:04:46 & -0.6& y & 2010\\
\hline
\end{tabular}
\tablefoot{
\tablefoottext{a}{http://www.sc.eso.org/santiago/uvespop/}
\tablefoottext{b}{Part of the spectrum was obtained within proposal P79.D-0810 and was kindly
provided for inclusion in the CRIRES-POP library.}
\tablefoottext{c}{No $M$-band spectrum available at the present time.}
}
\end{table*}

\subsection{Current content}
\subsubsection{\object{HD 149438} (\object{$\tau$\,Sco}), B0.2 V} 
At a distance of about 150\,pc, $\tau$\,Sco (spectral type B0.2\,V) is one of the brightest young massive stars of the southern Sco-Cen association. 
It is therefore one of the most intensely studied early-type stars, with observations 
reported from the X-ray regime to the radio. Notable data sets 
are the pioneering high-resolution spectral atlases obtained 
with both XMM-Newton in X-rays (Mewe et al. \cite{Meweetal03}) and with the Copernicus satellite in the UV (Rogerson 
\& Upson \cite{RoUp77}). Unusual stellar wind characteristics and unusually hard X-ray emission 
have been revealed at 
these short wavelengths despite the star's inconspicuousness in the optical spectrum. This is likely because of 
pronounced shocks arising from its clumped stellar wind (Howk et al. \cite{Howketal00}). Moreover, the star continues to offer 
observational surprises, such as the detection of its complex magnetic field by Donati et al.
(\cite{Donatietal06}). $\tau$\,Sco has been the subject of numerous quantitative model atmosphere analyses, starting with 
the first detailed spectral investigation of any star besides the Sun, by Uns\"old~(\cite{Unsoeld42}). A summary of 
stellar parameters determined in the more recent literature is given by Sim\'on-D\'iaz et al.
 (\cite{Simon-Diazetal06}) and Nieva \& Przybilla (\cite{np08}). 
 The most comprehensive non-LTE study to date found an effective temperature of $T_\mathrm{eff}$\,=\,
32\,000$\pm$300\,K, surface gravity $\log g$\,=\,4.30$\pm$0.05\,(cgs units), microturbulence $\xi$\,=\,
5$\pm$1\,km\,s$^{-1}$, projected equatorial rotational velocity $v \sin i$ and macroturbulence $\zeta$ of both 
4$\pm$4\,km\,s$^{-1}$, and elemental abundances typical of the massive star population in the solar neighbourhood 
(Przybilla et al. \cite{Przybillaetal08}), close to solar values. Exceptions are the light elements carbon and 
nitrogen, which indicate a high degree of mixing of the surface layers with CN-processed matter from the stellar core 
(Przybilla et al. \cite{Przybillaetal10}).

The first global spectral synthesis modeling for $\tau$\,Sco of an observed spectrum
of high quality in the optical was presented in Nieva \& Przybilla (\cite{np11}). The observed
spectrum was acquired with FEROS at high resolution (R=48\,000) and high
S/N ($\sim$800 in the $B$-band), and the global spectrum computed assuming non-LTE
covers the wavelength region from 3900 \AA to 6850 \AA.
A study of $\tau$\,Sco in the NIR shows very good agreement of parameter
determinations in comparison with the analysis in the optical. The He I
$\lambda $2.058 $\mu $m and $\lambda $2.11 $\mu $m obtained with
UKIRT/CGS4 (Mountain et al.\,\cite{mountain}) could be perfectly recovered with non-LTE techniques
(Nieva \& Przybilla \cite{np07}). Further quantitative analyses of CRIRES data collected
in 2007\footnote {The spectra of $\tau$\,Sco obtained in the run
079.D-0810(A) have been incorporated to the CRIRES-POP database by kind agreement of 
Norbert Przybilla.}
and corrected for telluric features following Seifahrt et al.~(\cite{Sei10}) found good
agreement between data and models several hydrogen and helium lines, as well
as metal lines, when adopting the stellar parameters and chemical
abundances derived in the optical (Nieva et al. \cite{nieva09}, \cite{nieva11}).

\subsubsection{\object{HD 47105} (\object{$\gamma$\,Gem}), A0 IV} 
Also known as Alhena, $\gamma$ Gem is a nearby A0\,IV star. Alhena has been found to be a wide binary from both
spectroscopic (Pourbaix et al. \cite{PTB04}) and astrometric (Jancart et al. \cite{JJBP05}) measurements, that has an orbital period of 4614.5\,d and an orbital
eccentricity of $e$ =\,0.89. The Hipparcos measurements (van Leeuwen \cite{Hipnew}) inferred 
a distance of 33.5\,pc with an uncertainty of
approximately 2.5\,pc. Its rotational velocity was 
determined to be 15\,km\,s$^{-1}$ (Royer et al. \cite{RZG07}), and its heliocentric radial
velocity $-$12.5\,km\,s$^{-1}$  (Valdes et al. \cite{VGRSB04}). Zorec et al. (\cite{zorec09}) derived 
$T_{\rm eff}$=9040$\pm$280\,K for this object
using the bolometric flux method.  Landstreet et al. (\cite{landstreet09}) used Str\"omgren and Geneva photometry to determine $T_{\rm eff}$=9150\,K and
$\log g$\,=\,3.5. From high resolution optical spectra, they measured  $v \sin i$\,=\,11$\pm$\,0.4\,km\,s$^{-1}$ with a microturbulence
$\xi$\,=\,1.2$\pm$0.4\,km\,s$^{-1}$. According to their analysis, the iron abundance is close to the solar value. This agrees with the results of
Adelman \& Unsuree (\cite{AU07}), who found [Fe/H]$=$0.11$\pm$0.18.  No UVES-POP spectrum is available for this star,  but it is part of the spectral
library produced by Valdes et al. (\cite{VGRSB04}) covering the wavelength range between 3465\,{\AA} 
and 9450\,{\AA} at a resolution of 1\,{\AA} FWHM. Flux-calibrated 
low-resolution $J$, $H$, and $K$ band spectra of this star were presented in the catalogue of  Ranade et al. (\cite{ranade07}).
\subsubsection{\object{HD 118022} (\object{$o$ Vir}), A1p}
o Vir (A1p) has the distinction of being the star for which
the pioneering work of Babcock (\cite{babcock47}) employed the Zeeman effect to determine
the magnetic field. Its magnetic field has since been well-studied 
(mean magnetic field $< B_s >$\,=\,3.0 kG, Ryabchickova et al. \cite{RKB08}),  in addition
to other atmospheric phenomena. An effective temperature of 9460\,K was
determined by Netopil et al. (\cite{NPMNH08}) in good agreement with the value used
in the analysis of Ryabchickova et al. (\cite{RKB08}). The latter also found $\log g$\,=\,4.0. 
Its light curve displays a small amplitude variation (0.07 mag in $V$),
and it is classified as an $\alpha^2$ CVn type variable, its variability being caused by the
rotation of an inhomogeneous stellar surface. Its measured value of 
$v \sin i$\ is 10\,km\,s$^{-1}$ (Ryabchikova et al.
\cite{RKB08}), which we verified by performing a synthetic spectrum analysis of UVES and CRIRES
data. The optical spectrum displays a richness caused by the rare earth
elements, which appears clearly evident at NIR wavelengths as broad but shallow
features. The presence of \ion{Eu}{ii} near 1 $\mu$m was noted by Wahlgren (\cite{wahlgren11}). The IR spectrum
will be utilised to identify atomic lines from heavy elements, and lines for magnetic field studies in chemically peculiar stars.

\subsubsection{\object{HD 73634} (\object{e Vel}), A6 II}
The bright giant e Vel (HR 3426) was proposed as
a low $v \sin i$ A6 II spectral standard by Gray \& Garrison (\cite{GrGa89}). 
The UVES-POP data illustrate the sharp-lined nature of the spectrum, along with several
broader, weak emission features at red wavelengths. From our synthetic spectrum fitting
of the UVES-POP spectrum,  $v \sin i$ was found to be 8.5$\pm$0.2 km\,s$^{-1}$.
Basic data derived from standard calibrations were presented by Snow et al.
(\cite{Snowetal94}), namely $T_{\rm eff}$\,=\,8100\,K, $M_{\rm bol}$\,=\,$-$2.8, 
and rv\,=\,19 km\,s$^{-1}$.

\subsubsection{\object{HD 20010} (\object{$\alpha$ For}), F8 V} %LHS 1515
Saffe et al. (\cite{Saffe08}) derived values for $T_{\rm eff}$ and 
$\log g$ of 6094\,K~(6226\,K) and 4.09 (3.77) dex, respectively 
(depending on the $T_{\rm eff}$-$\log g$-colour relation used). According 
to these values, the metallicity [Fe/H] is $-$0.64 ($-$0.62). 
A more comprehensive spectroscopic analysis by Gonzalez et al.
(\cite{Gonzalez10}) found $T_{\rm eff}$\,=\,6170$\pm$35\,K, $\log
g$\,=\,3.80$\pm$0.08 and [Fe/H]\,=\,$-$0.21$\pm$0.03, and
Bruntt et al. (\cite{Bruntt10}) derived $T_{\rm eff}$\,=\,6015\,K, 
$\log g$\,=\,3.80, 
$v \sin i$\,=\,3.9\,km\,s$^{-1}$, $\zeta$\,=\,3.7\,km\,s$^{-1}$, and
[Fe/H]\,=\,$-$0.28, with the latter value also being representative of
many more chemical species.
A parallax of 70.24\,mas was derived from 
Hipparcos observations (van Leeuwen \cite{Hipnew}), suggesting a 
distance of only 14\,pc.
Oudmaijer et al. (\cite{oud92}) searched for possible dust around this 
star using IRAS data, but found only a very small IR excess.

\subsubsection{\object{HD 109379} (\object{$\beta$ Crv}), G5 II}
The most recent determination of the parameters of this G5 II star was presented by Lyubmikov et al.
(\cite{LLRRP10}), giving $T_{\rm eff}$\,=\,5100$\pm$80\,K, $\log g$\,=\,2.52$\pm$0.03, and [Fe/H]\,$=$\,+0.15. 
Using the Hipparcos
parallax (van Leeuwen \cite{Hipnew}), a distance of 44.6 pc was derived, and a mass of 3.7$\pm$0.1\,$M_{\odot}$
determined. The study of Takeda et al. (\cite{TSM08}) found similar values of $T_{\rm eff}$\,=\,5145\,K, 
$\log g$\,=\,2.56, and [Fe/H]\,=\,$-$0.01, in addition to 
log~$L/L_{\odot}$\,=\,2.23 and M=3.31\,$M_{\odot}$. Older investigations
(e.g. Luck \& Wepfer \cite{LW95}, Brown et al. \cite{BSLD89}) came to similar conclusions. The heliocentric 
radial velocity was 
measured to be $-$8\,km\,s$^{-1}$ (e.g. Buscombe \& Morris \cite{BM58}), and $v \sin i$ 
was around 4\,km\,s$^{-1}$ (Gray \cite{Gray81}, Fekel \cite{Fekel97}). A radius of 11\,$R_{\odot}$
is listed in Pasinetti Fracassini et al. (\cite{pasin01}).

While the C/O ratio was 
found to be close to the solar value, the two elements C and O seem to be depleted in this
star relative to the solar value according to the analysis of Luck \& Wepfer (\cite{LW95}). 
An earlier claim of an enrichment in barium (Keenan \& Pitts \cite{KP80}) could not be confirmed (Eggen \& Iben \cite{EI91}).

\subsubsection{\object{HD 83240} (\object{10\,Leo}), K1 III}
Stellar parameters for this K1 giant, also known as the bright star HR\,3827 or 10\,Leo, were 
derived by Soubiran et 
al. (\cite{soubi08}). They found $T_{\rm eff}$ and $\log g$ to be 4682\,K and
2.45\,dex, respectively. Soubiran et al. evaluated a metallicity [Fe/H] of $-$0.02. Compared to Arcturus, 
which is thought 
to be a metal-poor object
(e.g. Peterson et al. \cite{peters93}), we thus have here a K giant reference spectrum of solar metallicity. The 
Hipparcos distance is 69\,pc.
The star is a spectroscopic binary with an orbital period of 2834$\pm$4\,d and $\gamma$=+20.0$\pm$0.1\,km\,s$^{-1}$ (Griffin 
\cite{griffi85}). No trace of
the secondary star was detected, the orbital parameters indicate that it is of low mass. Pasinetti Fracassini et 
al. (\cite{pasin01})
listed a radius of 14\,$R_{\odot}$ for this object. No indication of light variability was found (Percy 
\cite{percy93}).

Mishenina et al. (\cite{mishe06}) measured abundances of various key elements from the spectrum of HD 83240. They 
found that 
$\log A(\rm C)$\,=\,8.25, $\log A(\rm N)$\,=\,8.2, $\log A(\rm O)$\,=\,8.75, and $\log A(\rm Li)$\,=\,1.1. This means a slight underabundance in carbon and an overabundance in 
nitrogen, which agrees with a post first-dredge-up red giant. Silicon, calcium, and nickel are solar-like in abundance. 
In a second paper, Mishenina et al. (\cite{mishe07}) provided abundances for various neutron-capture elements. None of 
the element abundances differed from the solar value by more than the typical error given by the authors.

\subsubsection{\object{HD 225212}, K3 I}
The most recent and most detailed study of the global parameters of this K-type supergiant was 
provided by
Smiljanic et al. (\cite{SBDM06}), finding $T_{\rm eff}$\,=\,4052\,K, $\log g$\,=\,0.75, $\xi$\,=\,2.95\,km\,s$^{-1}$, and 
[Fe/H]\,=\,+0.1. They derived $M_{\rm bol}$\,=\,$-$4.42, and estimated a mass of this object of 7.0--7.9\,$M_{\odot}$. Their
heliocentric radial velocity value of $-$42\,km\,s$^{-1}$ is in excellent agreement with other measurements in the
literature (De Medeiros et al.\,\cite{DUBM02}, Valdes et al.\,\cite{VGRSB04}, Eaton \& Williamson \cite{EW07}).
Earlier studies of the stellar parameters found quite diverse results: van Paradijs (\cite{paradijs73})
measured a similar $T_{\rm eff}=$4100\,K, but a general overabundance of heavy elements by a factor of 1.6, while 
Luck \& Bond (\cite{LB80}) derived for almost the same effective temperature a metallicity of [Fe/H]\,=\,$-$0.2. This
scatter may result from the different values of 
the microturbulence used. Lower values for the temperature were
suggested by Blum et al. (\cite{BRSO03}, 3860\,K) and Mallik (\cite{mallik99}, 3689\,K). Values very similar
to the results by Smiljanic et al. (\cite{SBDM06}) can be found in the library of S{\'a}nchez-Bl{\'a}zquez et al.
(\cite{SPJ06}).

Smiljanic et al. derived abundances of [C/Fe]\,=\,$-$0.14, [N/Fe]\,=\,+0.41, and [O/Fe]\,=\,$-$0.02. 
These findings
agree with predictions for an atmospheric composition past first dredge-up for a non-rotating star. Consistently,
a rotational velocity of 5.8$\pm$1\,km\,s$^{-1}$ has been measured (De Medeiros et al. \cite{DUBM02}).

No indications of binarity have been detected (Eggleton \& Tokovinin \cite{ET08}). Koen \& Eyer (\cite{KE02})
found evidence for a two hour variability with an amplitude of a few mmag from the epoch photometry of the Hipparcos
catalogue. The parallax of this object is 1.56$\pm$0.31\,mas (van Leeuwen \cite{Hipnew}).

\subsubsection{\object{HD 49331}, M1 I}
Two studies were dedicated to the measurement of stellar parameters of this M supergiant. Smith \& Lambert 
(\cite{SL86}) derived $T_{\rm eff}$\,=\,3600\,K, $\log g$\,=\,0.7, and $M_{\rm bol}$\,=\,$-$3.9. The Hipparcos
distance leads to a somewhat brighter $M_{\rm bol}$\,=\,$-$4.5 (using the $V$ brightness and the bolometric correction
of Worthey \& Lee \cite{WL11}). Smith \& Lambert also determined the
abundances of various elements and found [Fe/H]\,=\,+0.14, $^{12}$C/$^{13}$C\,=\,18, and $^{12}$C/$^{16}$O\,=\,0.32,
with no indications for an enhancement in s-process elements. The more recent investigation of Wylie-De Boer \&
Cottrell (\cite{WC09}) confirmed temperature, $\log g$ value, and the 
slightly super-solar metallicity of HD 49331, but measured a higher
C/O ratio of 0.54 and a mild enhancement of s-process elements. The latter result is in agreement with an
early spectral classification of the star as MS due to enhanced bands of YO and ZrO (Yamashita \cite{yamashita}).
However, clear signs of technetium, a key indicator of recent third dredge-up, are missing (Little et al. \cite{LLMB}).

A mean radial velocity of 26.7\,km\,s$^{-1}$ was measured by Brown et al. (\cite{BSL90}).

\subsubsection{\object{HD 224935} (\object{YY Psc}), M3 III}
The bright, cool giant YY Psc (M3 III) has an accurate parallax determined by
the Hipparcos satellite of 7.55$\pm$0.59\,mas (distance of 132\,pc, van Leeuwen \cite{Hipnew}).
According to the GCVS (Samus et al. \cite{GCVS}), it displays irregular variability with an
amplitude of 0.1 mag at visual wavelengths, which was confirmed by
Hipparcos. Owing to the small amplitude of variability, 
we assumed that variability had no impact
on the analysis of the CRIRES spectrum. The angular diameter of YY Psc was
measured both via long baseline interferometry (Dyck et al. \cite{vvT98})
and by the lunar occultation technique (Fors et al. \cite{FRNP04}). Both methods
provided a value close to 7 mas. Dyck et al. also derived $T_{\rm eff}$ =
3647$\pm$184 K, and a similar temperature was found in the investigation
of Dumm \& Schild (\cite{DS98}), who listed a radius of 109 R$_{\sun}$ and a
mass of 2.5 M$_{\sun}$. Studies by Sloan \& Price (\cite{SP98}) and Kwok et al.~(\cite{KVB97})
found YY Psc to be dust-free.

Eaton \& Williamson (\cite{EW07}) measured a radial velocity of 12.05$\pm$0.43\,km s$^{-1}$
from a series of 103 observations. The rotation velocity, $v \sin i$, was determined
by Zamanov et al. (\cite{Zamanov08}) to be either 2.5 or 3.9 km s$^{-1}$, depending on
the method applied. Lazaro et al. (\cite{Lazaro91}) measured a $^{12}$C/$^{13}$C
ratio of ten from low dispersion IR spectra, with a carbon abundance
that is deficient by [C/H]\,=\,$-$1.25, relative to the Sun. Using the CRIRES-POP
spectrum for a preliminary synthetic spectrum analysis with a model
of the parameters $T_{\rm eff}$\,=\,3500 K and $\log g$\,=\,1.0 (cgs),
Wahlgren et al. (\cite{WLW11}) confirmed the $^{12}$C/$^{13}$C
value of Lazaro, but not the carbon abundance, which was found to be
near solar in value. The abundances of both nitrogen and
oxygen (both near solar) were also determined and enhancements for the heavy elements
(Co, Ge, Sr, Cs, Ba, and Eu). Lines of Si, Y, and Zr were also identified in the
CRIRES spectrum. The abundance of iron is determined to be solar. 
The value for $v \sin i$ is 4\,km\,s$^{-1}$.

\subsubsection{\object{Barnard's star} (\object{V2500 Oph}, \object{GJ 699}), M4 V} Probably the most prominent star in the CRIRES-POP library is Barnard's star (Barnard \cite{barnard06}), the
star with the largest known proper motion. It is the brightest known M dwarf and the only observable M dwarf within our CRIRES-POP brightness limits.
Stellar parameters found in the literature for this star agree very well. The most recent study of Chavez \& Lambert (\cite{CL09}) measured $T_{\rm eff}$=3134\,K, $\log g$\,=\,5.1, $\xi$\,=\,0.6 km\,s$^{-1}$, 
[Fe/H]\,=\,$-$0.8, and [Ti/H]\,=\,$-$0.61. Using flux ratios from various wavelength bands, Casagrande et al.
(\cite{CFB08}) found $T_{\rm eff}$\,=\,3145$\pm$69\,K, and  S\'egranasan et al. (\cite{SKFQ03}) derived $T_{\rm eff}$\,=\,3163$\pm$65\,K, $\log g$\,=\,5.05$\pm$0.09, 
$M$\,=\,0.158$\pm$0.008\,$M_{\odot}$, and $R$\,=\,0.196$\pm$0.008\,$R_{\odot}$ using VLTI measurements. Dawson \& De Robertis (\cite{DD04}) determined a
luminosity of $L$\,=\,(3.46$\pm$0.17$)\cdot$10$^{-3}$\,$L_{\odot}$ and  $T_{\rm eff}$\,=\,3134$\pm$102\,K. 

The star's radial velocity was found to be $-$110.5\,km\,s$^{-1}$ (Nidever et al. \cite{NMBFV02}) with a very high degree of stability at the
3-5\,m\,s$^{-1}$ level (e.g. Bean et al. \cite{BSH10}).  Owing to its proximity and large space motion, it shows a non-negligible amount of secular
acceleration of  approximately 4.5\,m\,s$^{-1}$\,yr$^{-1}$ (Kuerster et al. \cite{kuerst03}). Benedict et al. (\cite{bene98})  measured a rotation period
of approximately 130 days for this star. The rotational velocity is thus very small and only upper limits have been reported (e.g.,
$v \sin i\ll$2.8\,km\,s$^{-1}$, Mohanty \& Basri \cite{MB03}). This old object (7--13 Gyr, Feltzing \& Bensby \cite{FB08}) shows only a low level of
magnetic activity (e.g. H\"ubsch et al. \cite{HSSV99}), although 
Paulson et al. (\cite{PAA06}) investigated a rare flare event on this star.  

\subsubsection{\object{HD 61913},  (\object{NZ Gem}), S}
The spectral classification of this star is slightly unclear. 
It was classified as M3S by Keenan (\cite{keenan54}). Yamashita (\cite{yamashita}) found that ZrO
and YO were 
slightly enhanced, while Smith \& Lambert (\cite{SL88}) found no enhancement in s-process elements.
Nevertheless, 
the star is listed in several later papers among MS/S stars, and our own inspection of the 
UVES-POP spectrum indicates a spectral type of MS owing to the clearly visible ZrO bands.
Famaey et al. (\cite{famaey}) summarized data about this star's distance and kinematics, namely d$=$324\,pc and
rv$_{hel}$=$-$15.2\,km\,s$^{-1}$. This velocity is in good agreement with findings from Jorissen et al. (\cite{Jori98}).
The latter authors also found some velocity jitter in this star, but no clear indication for binarity. 
Van Eck et al. (\cite{vanEck}) derived an absolute bolometric magnitude of $-$4\fm01.
Cenarro et al. (\cite{cenarro}) calculated $T_{\rm eff}=$3530\,K and $\log g$\,=\,0.7 from the spectral type
and luminosity class (M3II-III).
NZ Gem shows no lines of technetium (Lebzelter \& Hron \cite{LH03}), which suggests that any possible 
enhancement in s-process elements is due to an extrinsic enrichment. However, no indications for 
a binarity in this object were found in the study of Eggleton \& Tokovinin (\cite{ET08}). 

\subsubsection{\object{HD 134453} (\object{X TrA}), C}
The group of C-rich stars is represented in our library by the bright star HR 5644 (X TrA).  
A spectral type 
C5,5 was 
assigned to this object (Wallerstein \& Knapp \cite{WK98}). The star is an irregular variable with a 
maximum amplitude of 1 magnitude in the visual according to the GCVS (Samus et al. \cite{GCVS}). Percy et al. 
(\cite{Percy09}) estimated a variability period of 500 days. In a multiplicity study of bright stars by 
Eggleton \& Tokovinin (\cite{ET08}), the star is listed as a single object. A distance of 460 pc, based on the 
Hipparcos parallax, is given by Winters et al. (\cite{WLJNE03}). These authors also 
list a LSR-velocity for X TrA of 
$-$2.5\,km\,s$^{-1}$, corresponding to a heliocentric velocity of +2.5\,km\,s$^{-1}$. This is in agreement with 
measurements by Sch\"oier \& Olofsson (\cite{SO01}).

Kipper (\cite{Kipper04}) summarized earlier estimates of the star's effective temperature
and absolute bolometric magnitude, which he used to derive surface gravity and microturbulence from the 
UVES-POP spectrum. He determined the values of $T_{\rm eff}$\,=\,2700\,K, $M_{\rm bol}$\,=\,$-$6.0, 
$\log g$\,=\,$-$0.8, and 
$\xi$\,=\,2.3\,km\,s$^{-1}$, respectively. According to Kipper's analysis, X TrA seems to be slightly 
metal-poor 
([Fe/H]$=-$0.5) with a supersolar abundance of the s-process elements Y and Zr. The carbon isotopic ratio 
$^{12}$C/$^{13}$C was found to be 31. A C/O ratio of 1.1 was used for all carbon stars investigated by Kipper. The 
star probably displays 
lines of Tc (Little et al.\,\cite{LLMB}), which we confirmed by inspecting the UVES-POP
spectrum. Together with enhanced abundances of Y and Zr, this 
suggests that the star is an intrinsic AGB star, although the carbon isotopic ratio reported in the literature seems 
a bit low. Winters et al. (\cite{WLJNE03}) derived a mass loss rate of about 
5$\times$10$^{-7}$ M$_{\odot}$yr$^{-1}$ from thermal CO lines.

\subsection{Forthcoming targets}
The CRIRES-POP observations are expected to continue for an additional year to reach a total sample size of approximately
25-30 targets. Spectra have been partly obtained for seven further targets, which are listed in
Table \ref{future}. Information about the stars
will be published online\footnote{http://www.univie.ac.at/crirespop} 
as soon as observations are completed. 

\begin{table*}
\caption{Forthcoming additions to the CRIRES-POP library, 
sorted by spectral type.}
\label{future}
\centering
\begin{tabular}{llcrrcc}
\hline\hline
HD number & other names & spectral type & RA (2000) & DE (2000) & $K$  & UVES-POP?\\
          &             &               &           &           & (mag) &\\
\hline
120709 & 3 Cen A, \object{HR 5210} & B5 IIIp & 13:51:50 & $-$32:50:39 & 4.97 &  y\\
80404 & $\iota$ Car, \object{HR 3699} & A8 I & 09:17:05 & $-$59:16:31 & 1.53 & y\\
146233 & 18 Sco, \object{HR 6060} & G2 V & 16:15:37 & $-$08:22:10 & 4.19 & n\\
99648 & $\tau$ Leo, \object{HR 4418} & G8 I & 11:27:56 & +02:51:23 & 2.83 & y\\
138716 & 37 Lib, \object{HR 5777} & K1 V & 15:34:11 & $-$10:03:52 & 2.24 & y\\
209100 & $\epsilon$ Ind, \object{HR 8387} & K5 V & 22:03:22 & $-$56:47:09 & 2.20 & y\\
73739 & \object{MN Vel} & M7 II/III & 08:38:01 & $-$46:54:15 & $-$0.19 & n\\
\hline
\end{tabular}
\end{table*}

\subsection{The CRIRES-POP HRD}
\begin{figure}
  \resizebox{\hsize}{!}{\includegraphics{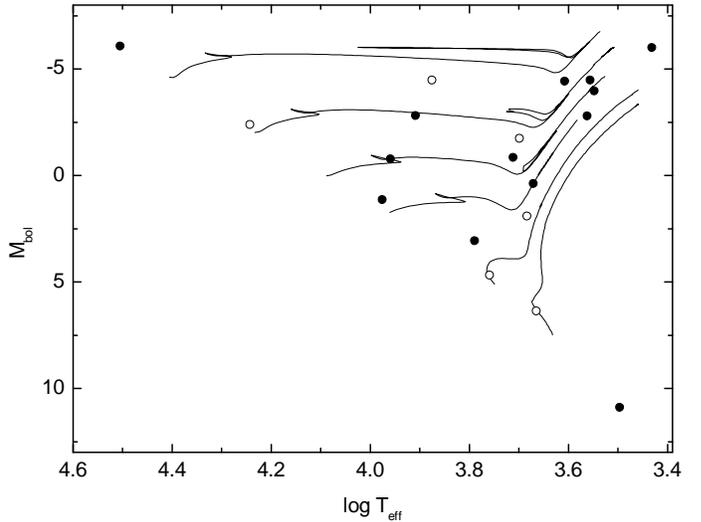}}
  \caption{Location of the CRIRES-POP stars in a HRD. Parameters taken from the literature or derived as 
  described in the text. Filled symbols mark stars presented here, and open symbols the forthcoming objects.
  The solid lines indicate evolutionary tracks for solar-metallicity models of 10, 5, 3, 2, 1, and 0.6 M$_{\odot}$ 
  taken from Bertelli et al. (\cite{bertelli08}, \cite{bertelli09}).}
  \label{pop-hrd}
\end{figure}

We include in Fig.\,\ref{pop-hrd} 
the location of the stars of the CRIRES-POP library in a Hertzsprung-Russell diagram
(HRD). Temperatures and luminosities were taken from the literature, and for each source 
we refer to the individual target descriptions in the paragraphs above. If no luminosity had previously been calculated, we used the Hipparcos parallax
(van Leeuwen \cite{Hipnew}), temperature, colour, and $V$ magnitude from the literature as described above, and the bolometric correction from Worthey \& Lee
(\cite{WL11}) to estimate a value for $M_{\rm bol}$. Luminosities from the literature were converted into 
$M_{\rm bol}$ using $M_{\rm bol, \odot}$\,=\,4$\fm$74. In Fig.\,\ref{pop-hrd} we did not included 
the star
HD 73739 owing to a lack of sufficient information about this object in the literature.

We include in Fig.\,\ref{pop-hrd} evolutionary tracks 
for stars with solar metallicity and masses of 10, 5, 3, 2, 1, and 0.6 M$_{\odot}$, taken from
Bertelli et al. (\cite{bertelli08}, \cite{bertelli09}). The tracks are primarily for illustrative purposes.
Owing 
to the inhomogeneous collection of temperature and luminosity values presented here, we did 
not aim to
perform fits of the mass and evolutionary status of the various targets. Our intention was instead to 
highlight the parameter range covered by CRIRES-POP.

\section{Results and discussion}
\subsection{A first glance at the catalogue}
\begin{figure}
  \resizebox{\hsize}{!}{\includegraphics{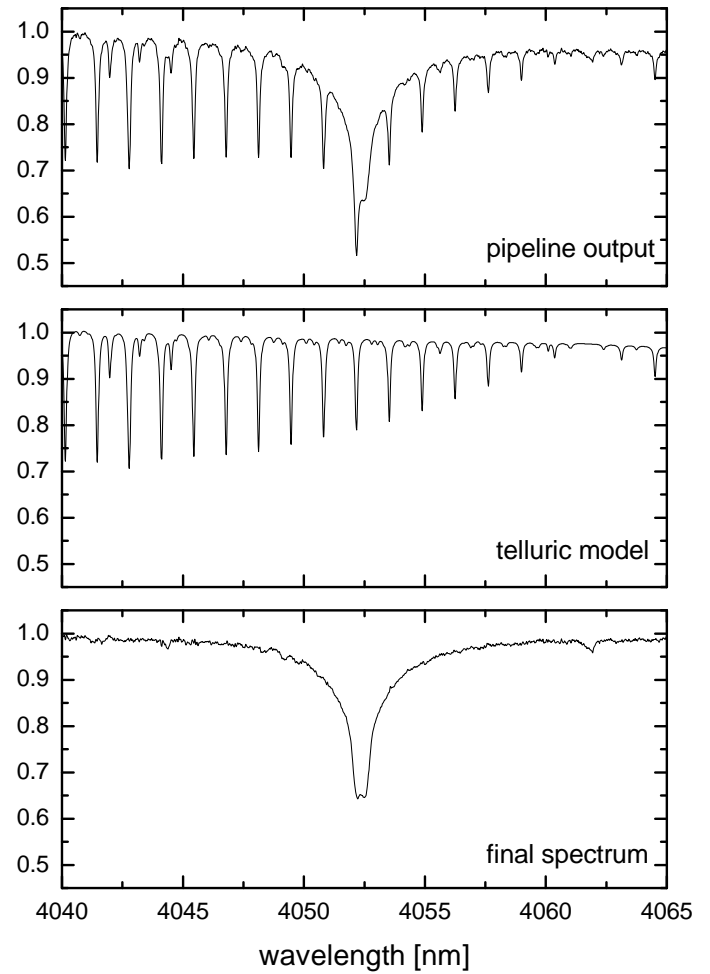}}
  \caption{ Example of our method of correcting for the telluric absorption. The top panel shows the spectrum 
  of HD 73634 (A6 II) around the Br-$\alpha$ line at the 
  level of the CRIRES pipeline data product after merging of two slightly overlapping spectral regions. The 
  middle panel shows the telluric model calculated to fit the telluric lines in the observed spectrum. The bottom 
  panel shows the resulting spectrum after division by the telluric spectrum and the application of some basic 
  cosmetic corrections. Note that the
emission core in the broad hydrogen line is real and not a residual of a
telluric line.}
  \label{telcor}
\end{figure}

As outlined above, the final CRIRES-POP data products should clearly have an accuracy 
that exceeds the average level of the standard CRIRES pipeline products.
In particular, the wavelength solution has been improved by the usage of 
telluric lines (see Sect. 3.2), cosmetic defects have been corrected where possible
(see Sect. 2), and the full spectral resolution with a 
resolving power of $R \approx 100\,000$ will be maintained,
compared to $R \approx 80\,000$--90\,000 from the standard pipeline reduction (see Sect. 3.1).
We started the production of the final library, and presented here
the first examples. In Fig.\,\ref{telcor}, we illustrate the process of telluric correction and fine tuning
of the wavelength calibration briefly described in Sect.\,\ref{telluric}. This method has been applied in three 
representative wavelength regions for the 13 stars listed in Table\,\ref{t:sample}. We have 
selected these
parts of the spectra because they include the positions of some well-known hydrogen lines, several atomic lines,
and a large collection of molecular lines, i.e., interesting features that 
can be seen in all stars of our library
in this presentation.

\begin{landscape}
\begin{figure}
%\begin{sidewaysfigure*}
\centering
   \includegraphics[width=1.52\textwidth]{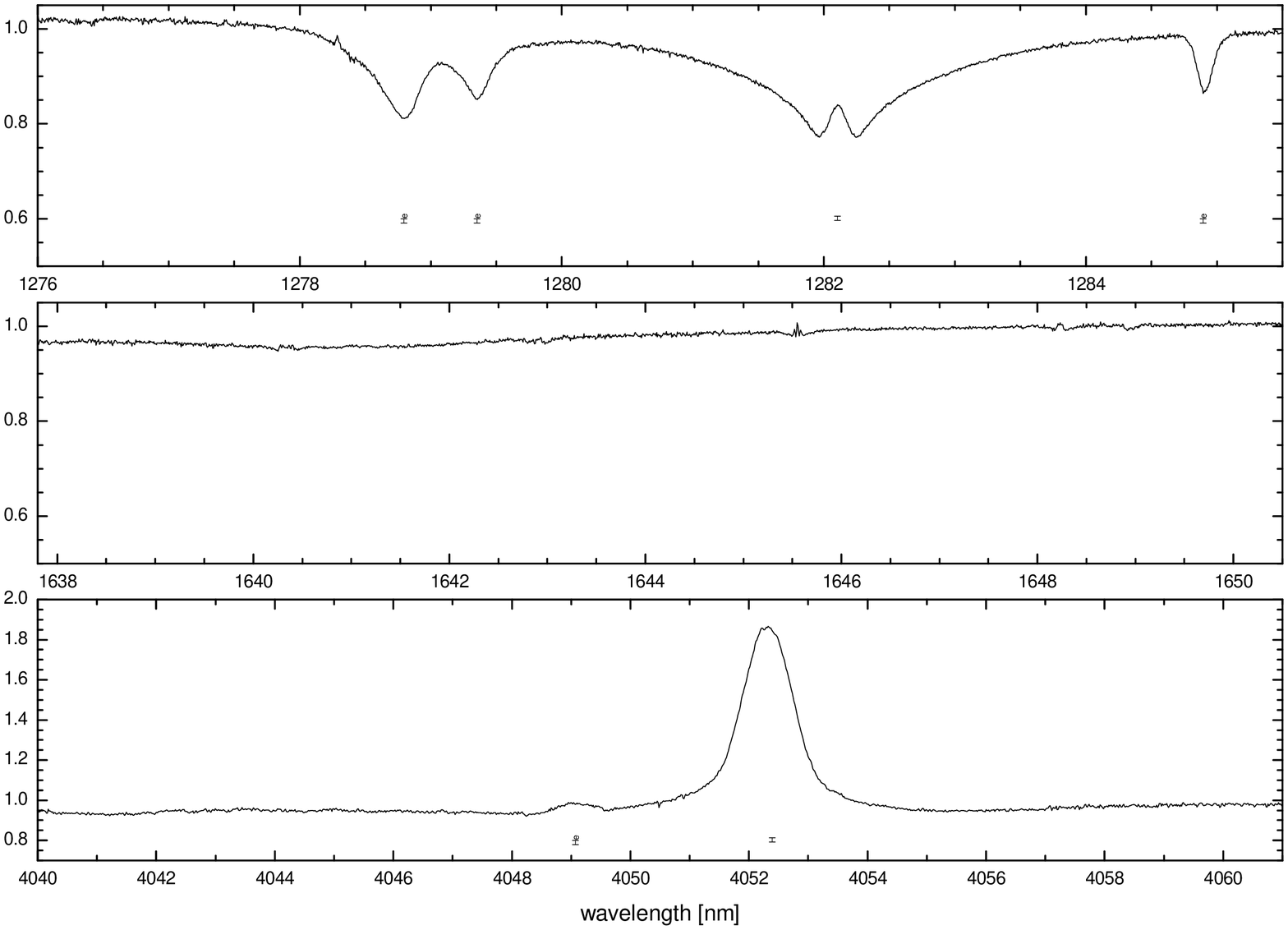}
\centering   
   \caption{Three representative sections of the CRIRES-POP data of HD 149438 (B0.2 V). 
   Telluric lines have been removed as described in the text. Known atomic and 
   molecular features are marked.}
\label{tausco}
\end{figure}
\end{landscape}

\begin{landscape}
\begin{figure}
\centering
   \includegraphics[width=1.52\textwidth]{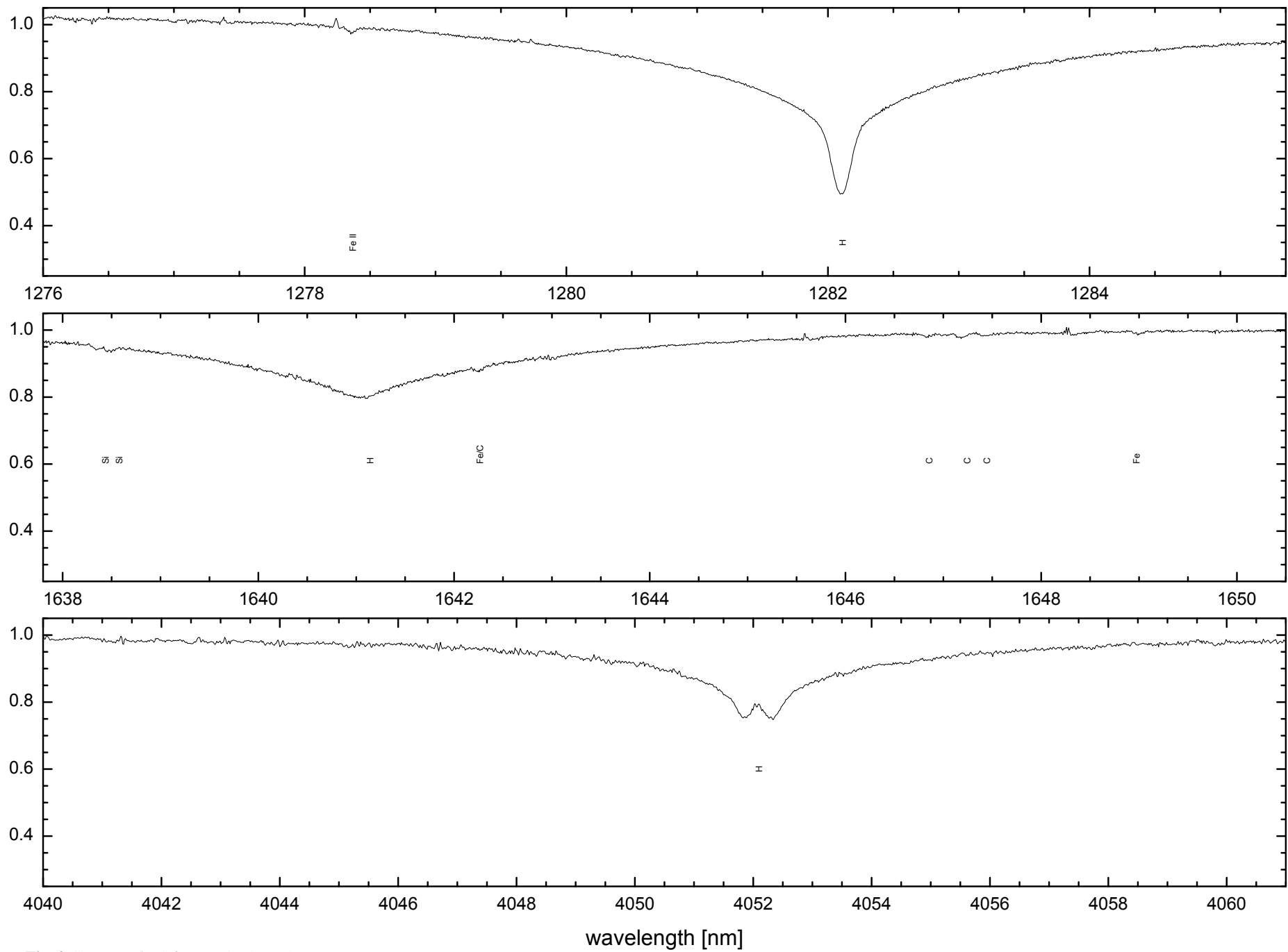}
\centering   
   \caption{Same as Fig.\,\ref{tausco} for HD 47105 (A0 IV).}
\label{gamgem}
\end{figure}
\end{landscape}

\begin{landscape}
\begin{figure}
\centering
   \includegraphics[width=1.52\textwidth]{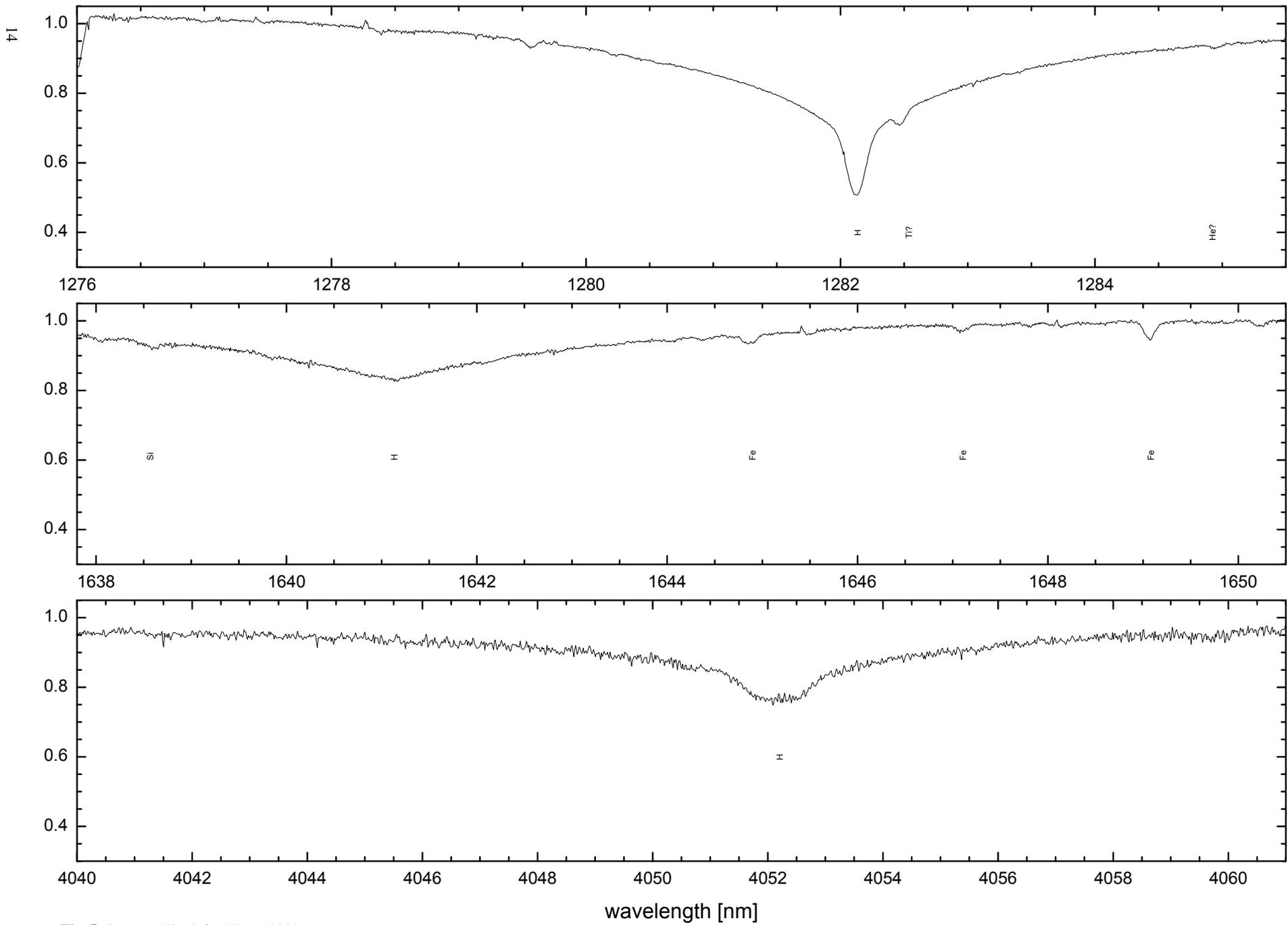}
\centering   
   \caption{Same as Fig.\,\ref{tausco} for HD 118022 (A1p).}
\label{ovir}
\end{figure}
\end{landscape}

Figures\,\ref{tausco} to \ref{xtra} show the spectra of the three wavelength regions for each star in the library and provide line
identifications. In our presentation, we follow the sorting given in Table\,\ref{t:sample}. 
Heliocentric velocity corrections have been applied to all spectra. 
The accuracy of the wavelength solution, derived with the help of the positions of 
the telluric lines, is better than one third of a pixel, and 
for the hot stars the accuracy is close to one tenth of a
pixel. A representative synthetic
telluric spectrum for these regions can be found in Fig.\,\ref{telmod}.
The telluric corrections applied to the
selected spectra shown here leave residuals in the cores of the most
prominent telluric lines of $\sim$1\% of the continuum flux for all but one
noticeable exception at 1278.5\,nm. The corrections were achieved without
back-fitting of line parameters and using a single Gaussian instrumental
profile.
The figures clearly illustrate the high quality of the CRIRES-POP data. 

\subsection{Line identifications} 
\begin{landscape}
\begin{figure}
\centering
   \includegraphics[width=1.52\textwidth]{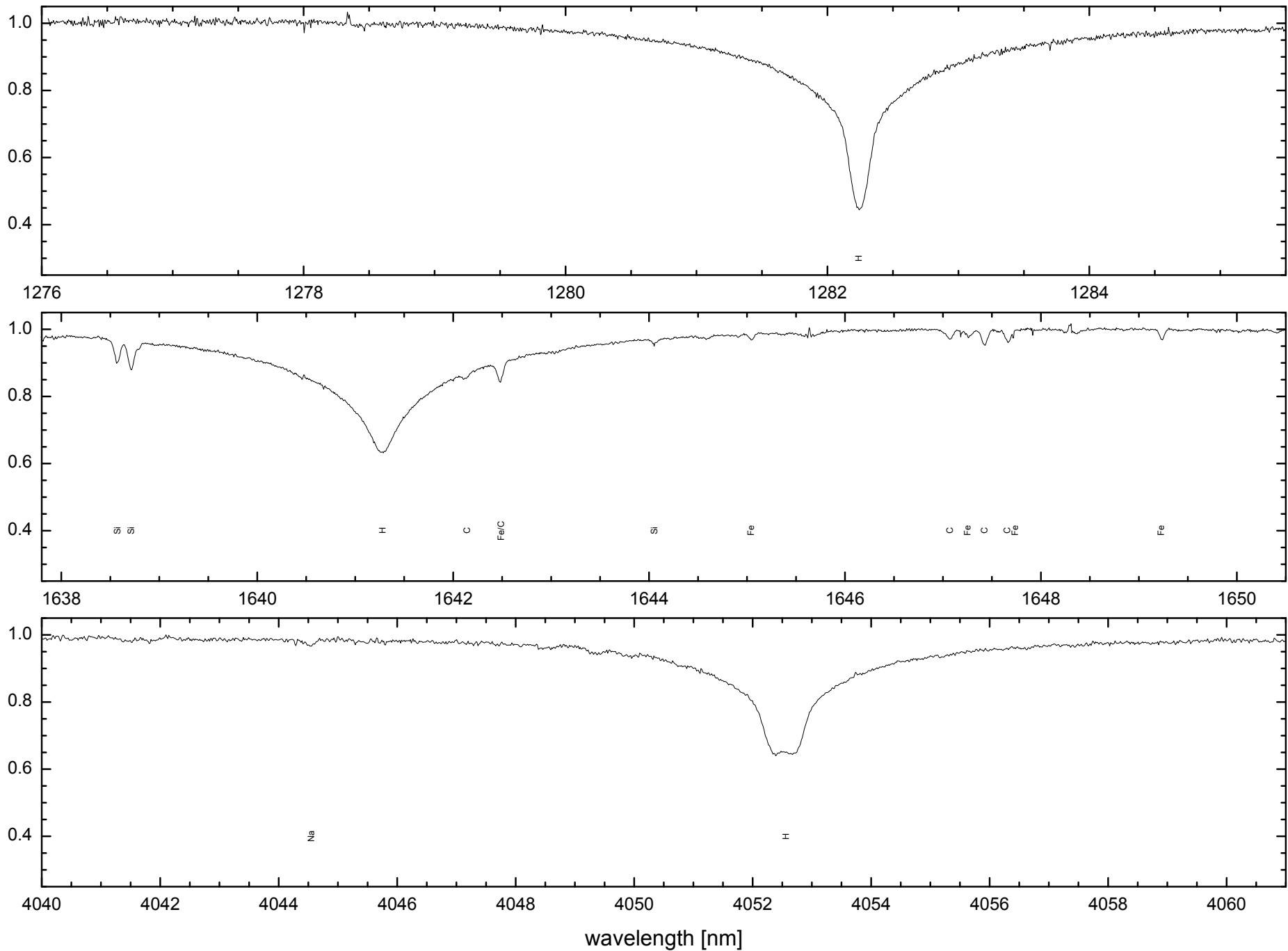}
\centering   
   \caption{Same as Fig.\,\ref{tausco} for  HD 73634 (A6 II).} 
\label{evel}
\end{figure}
\end{landscape}

\begin{landscape}
\begin{figure}
\centering
   \includegraphics[width=1.52\textwidth]{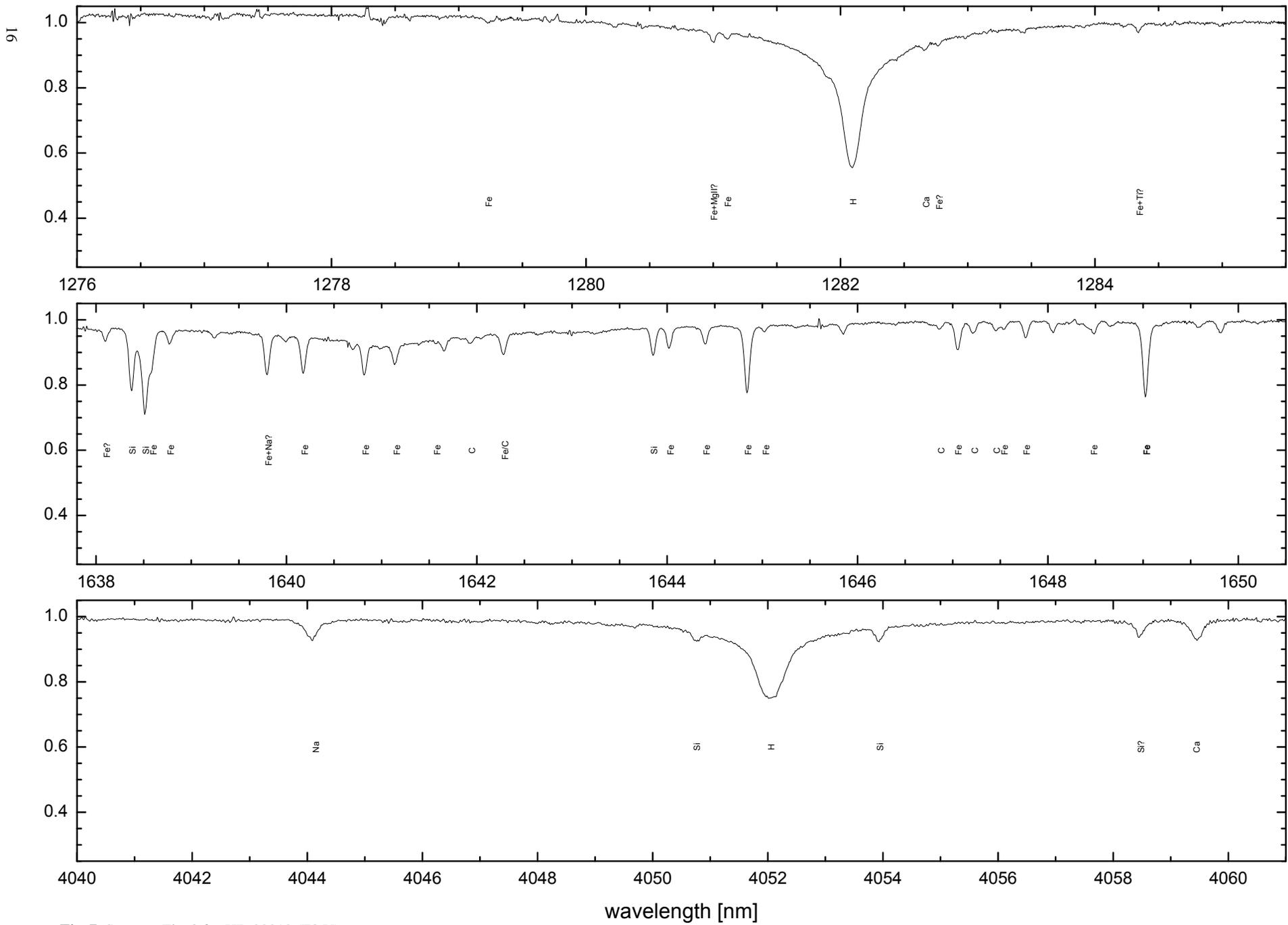}
\centering   
   \caption{Same as Fig.\,\ref{tausco} for HD 20010 (F8 V). }
\label{lhs1515}
\end{figure}
\end{landscape}

\begin{landscape}
\begin{figure}
\centering
   \includegraphics[width=1.52\textwidth]{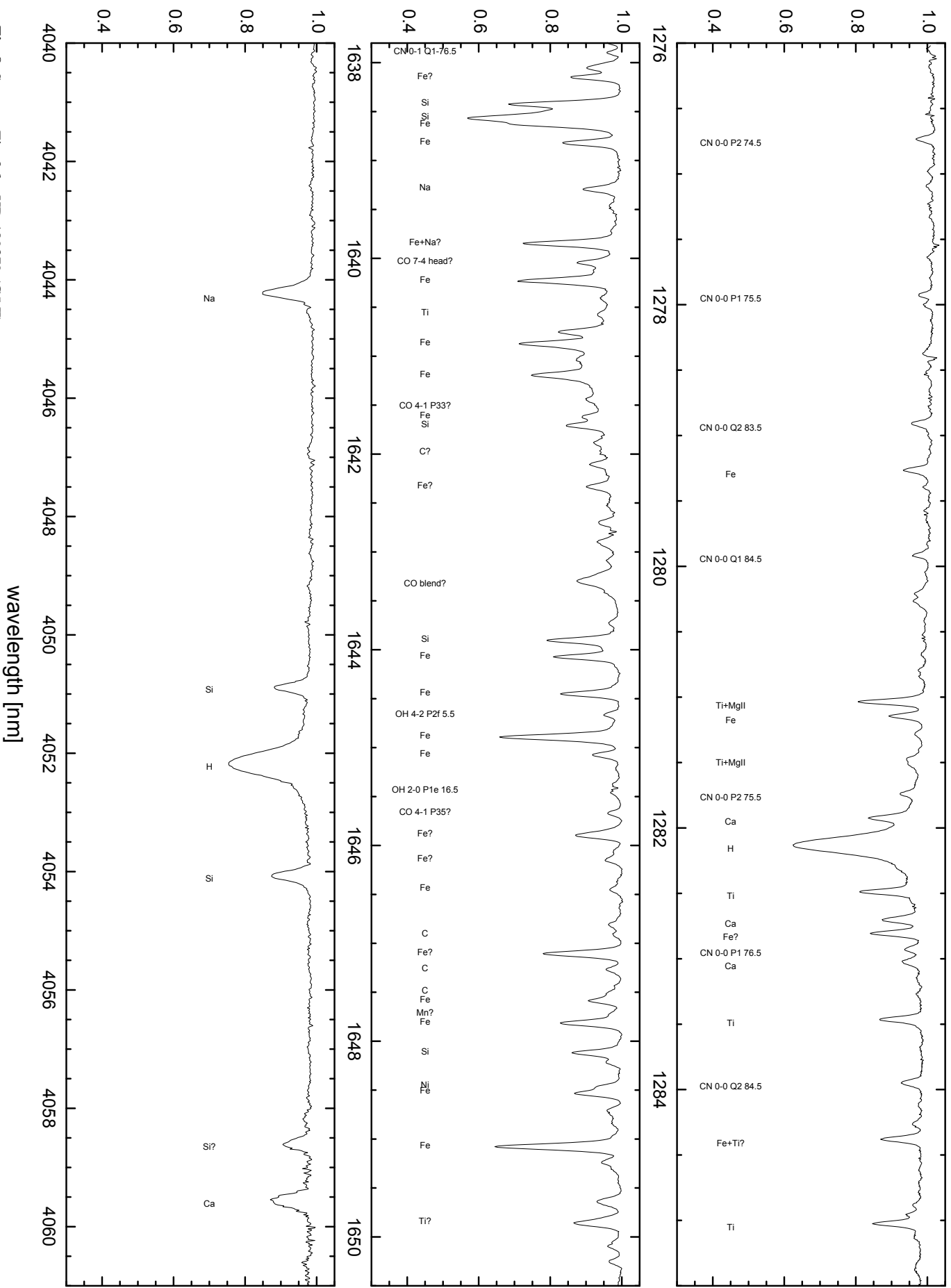}
\centering   
   \caption{Same as Fig.\,\ref{tausco} for HD 109379 (G5 II).} 
\label{hd109379}
\end{figure}
\end{landscape}

\begin{landscape}
\begin{figure}
\centering
   \includegraphics[width=1.52\textwidth]{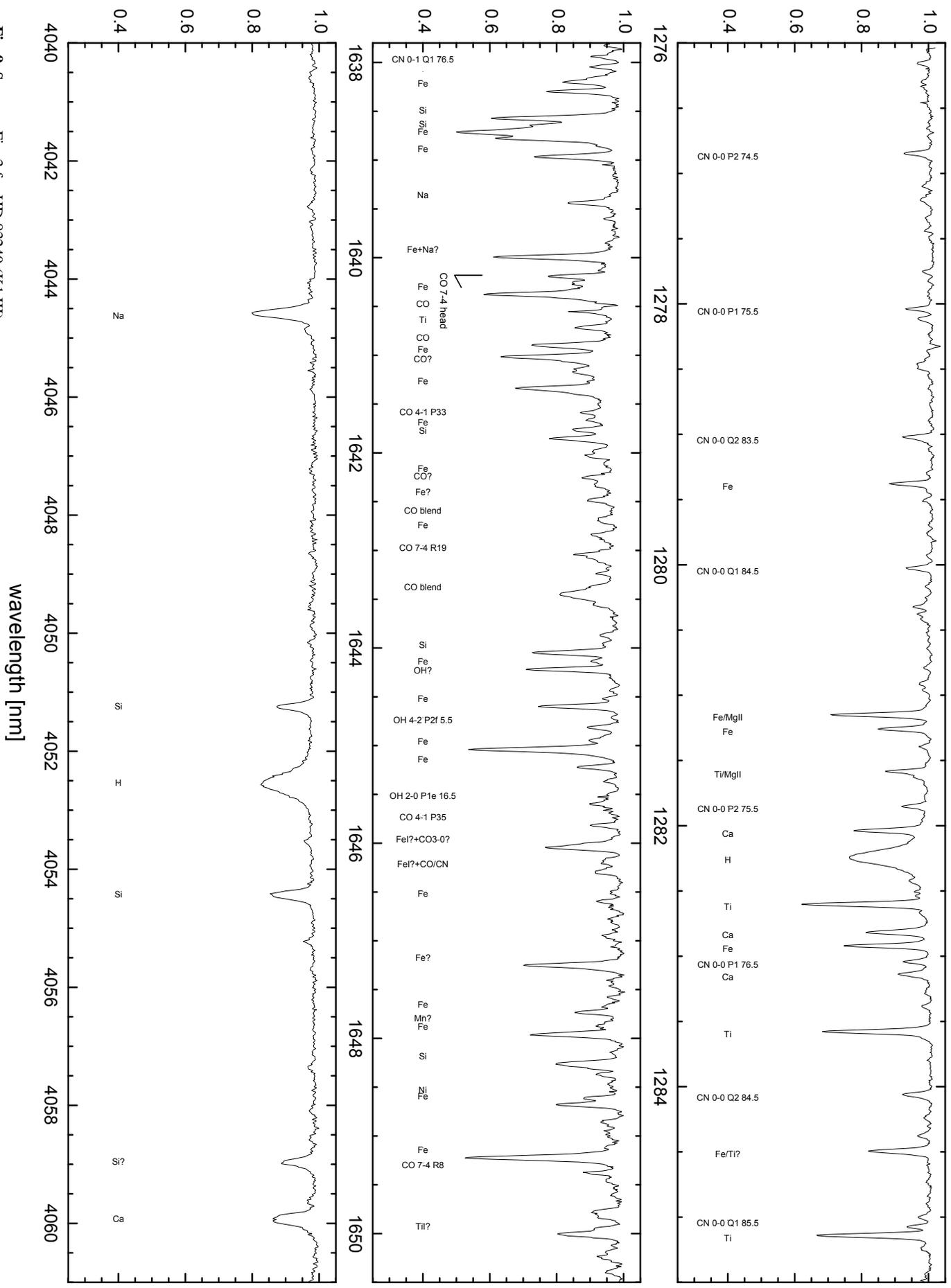}
\centering   
   \caption{Same as Fig.\,\ref{tausco} for HD 83240 (K1 III). }
\label{hd83240}
\end{figure}
\end{landscape}

\begin{landscape}
\begin{figure}
\centering
   \includegraphics[width=1.52\textwidth]{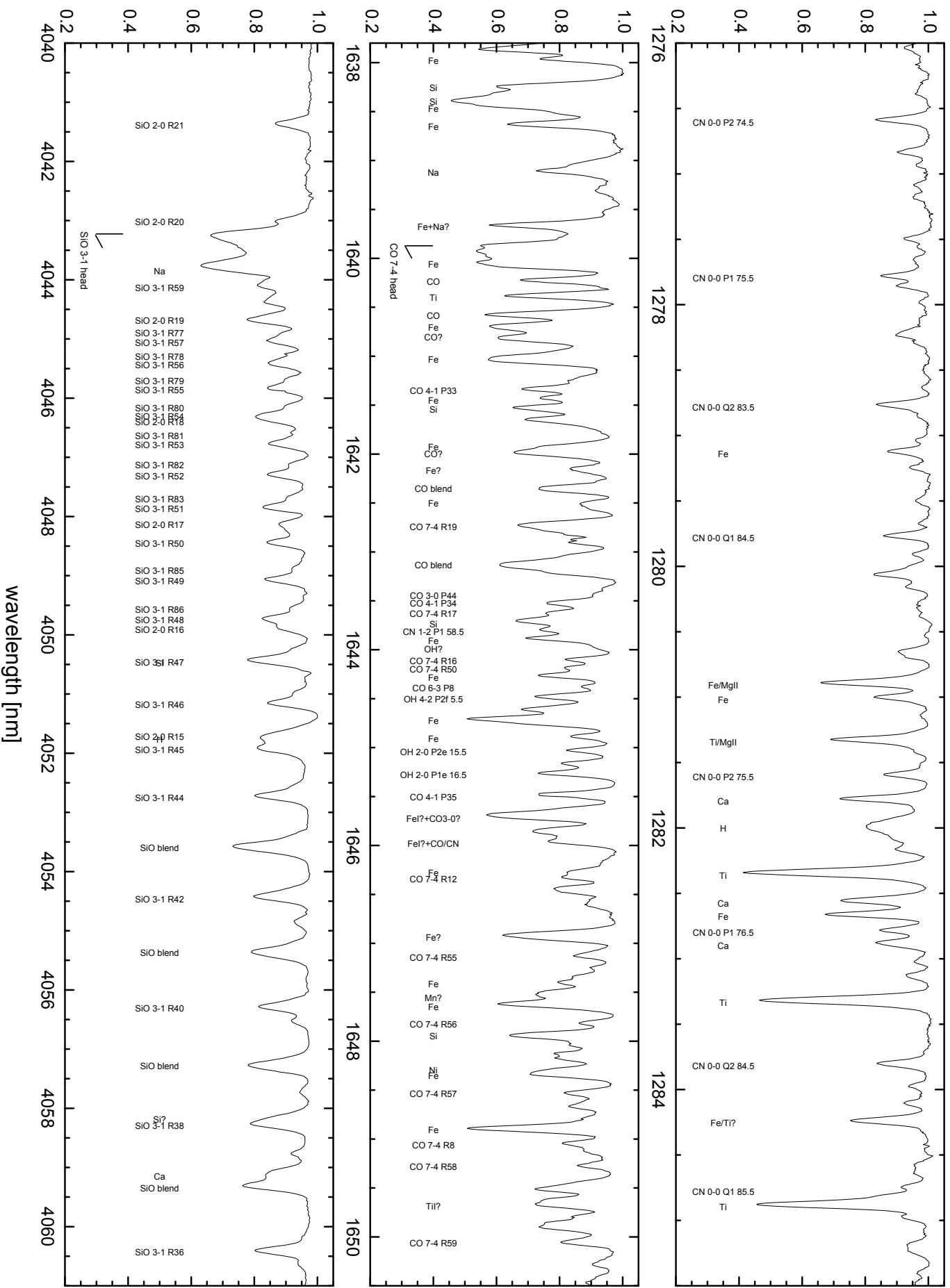}
\centering   
   \caption{Same as Fig.\,\ref{tausco} for HD 225212 (K3 I). }
\label{hd225212}
\end{figure}
\end{landscape}

\begin{landscape}
\begin{figure}
\centering
   \includegraphics[width=1.52\textwidth]{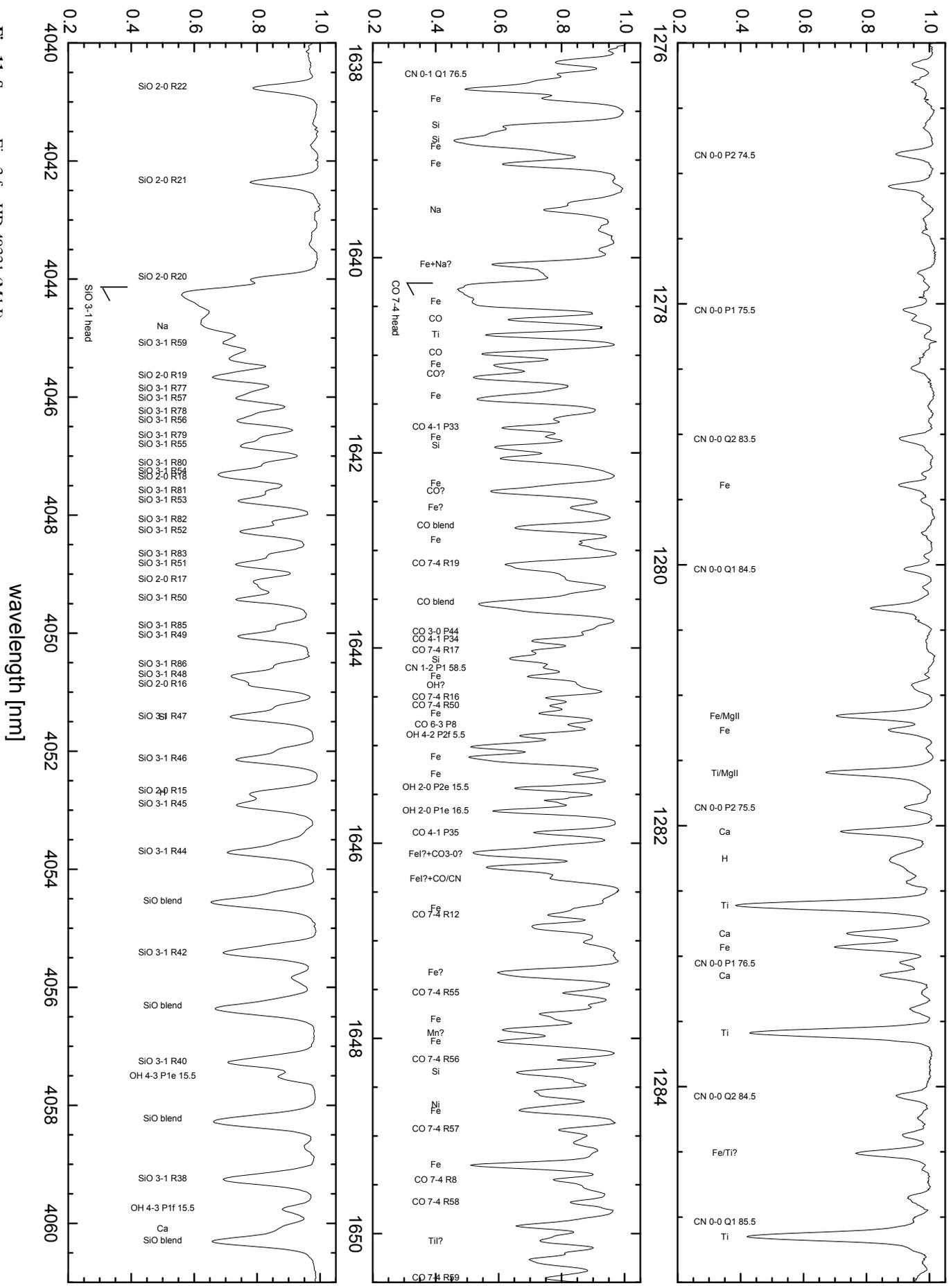}
\centering   
   \caption{Same as Fig.\,\ref{tausco} for HD 49331 (M1 I). }
\label{hd49331}
\end{figure}
\end{landscape}

\begin{landscape}
\begin{figure}
\centering
   \includegraphics[width=1.52\textwidth]{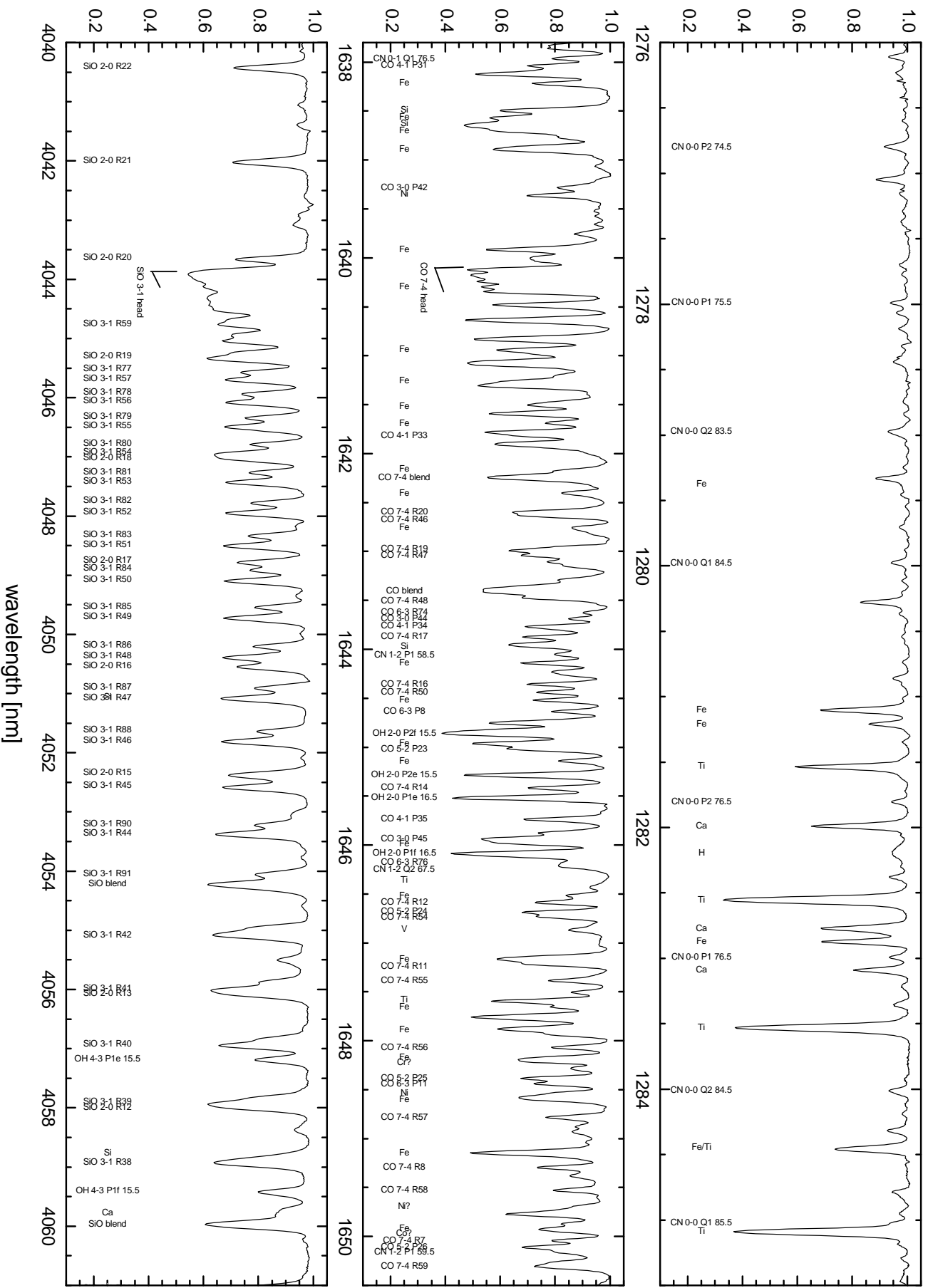}
\centering   
   \caption{Same as Fig.\,\ref{tausco} for HD 224935 (M3 III). }
\label{yypsc}
\end{figure}
\end{landscape}

In the presented spectra, atomic and molecular features have been identified. We 
used the identifications given in the
infrared atlas of Arcturus (Hinkle et al. \cite{arcturus}), the line lists provided by Kurucz \& Bell (\cite{kurucz95}),
the list of iron lines in Nave et al.
(\cite{nave94}), the HITRAN database from Rothman et al. (\cite{Roth08}), 
and the identifications of SiO and OH 
lines at 4\,$\mu$m given in Ridgway et al.
(\cite{ridgway84}). The attribution of atomic or molecular transitions  to the observed features was based on 
identifications for similar stars presented
in the literature or the most probable origin of the line, and have in some cases been verified by spectral 
modelling. One of the weakest lines identified is probably a Na line at 4044.5\,nm in HD 73634 (Fig.\,\ref{evel}), which has a central depth of not more than 2\,\%. Unidentified lines even weaker than that can be seen
e.g. in the $H$-band spectrum of the same star.

Line identifications for
$\tau$ Sco were taken from Nieva et al. (in preparation). 
Non-LTE population inversions in the outer atmosphere give rise to
emission cores in Br\,$\alpha$ in the hotter stars (see, e.g., K\"aufl \cite{Kaufl93},
Przybilla et al.~\cite{PB04}), 
that are most notable for $\tau$\,Sco, Pa\,$\beta$ also
displaying the phenomenon.

Line identification in the spectra of the majority of the observed targets was straightforwardly achieved; however, the identifications for two stars (the
M dwarf, Barnard's star, and the carbon star X TrA) required a special treatment owing to the lack of information available in the literature for these
spectral types. For these two stars, the spectra and line identifications are discussed in detail below.

\subsubsection{Barnard's star (GJ 699)}

The closest high-resolution template to our spectrum of Barnard's star (GJ\,699), is
the FTS atlas spectrum of a sunspot (i.e. dark umbra) acquired 
by Wallace et al. (\cite{WHL01}, \cite{WHL02}). 
However, the available spectra of the sunspot and GJ\,699 differ remarkably, which is not 
completely surprising, given the high magnetic field
and hence low gas pressure in a sunspot (e.g., Solanki \cite{solanki03}), mimicking a
low surface gravity environment (Amado et al. \cite{amado00}). In addition, GJ\,699 has a
lower metallicity ([Fe/H]\,=\,$-$0.5, Jones et al. \cite{JPVT02})
compared with the Sun. 

Moreover, the high magnetic field in the sunspot leads to strong Zeeman splitting and 
broadening of atomic (e.g. Ti~I and Ca~I around
1250\,nm) and molecular (FeH) lines, which we did not detect in GJ\,699.

\begin{landscape}
\begin{figure}
\centering
   \includegraphics[width=1.52\textwidth]{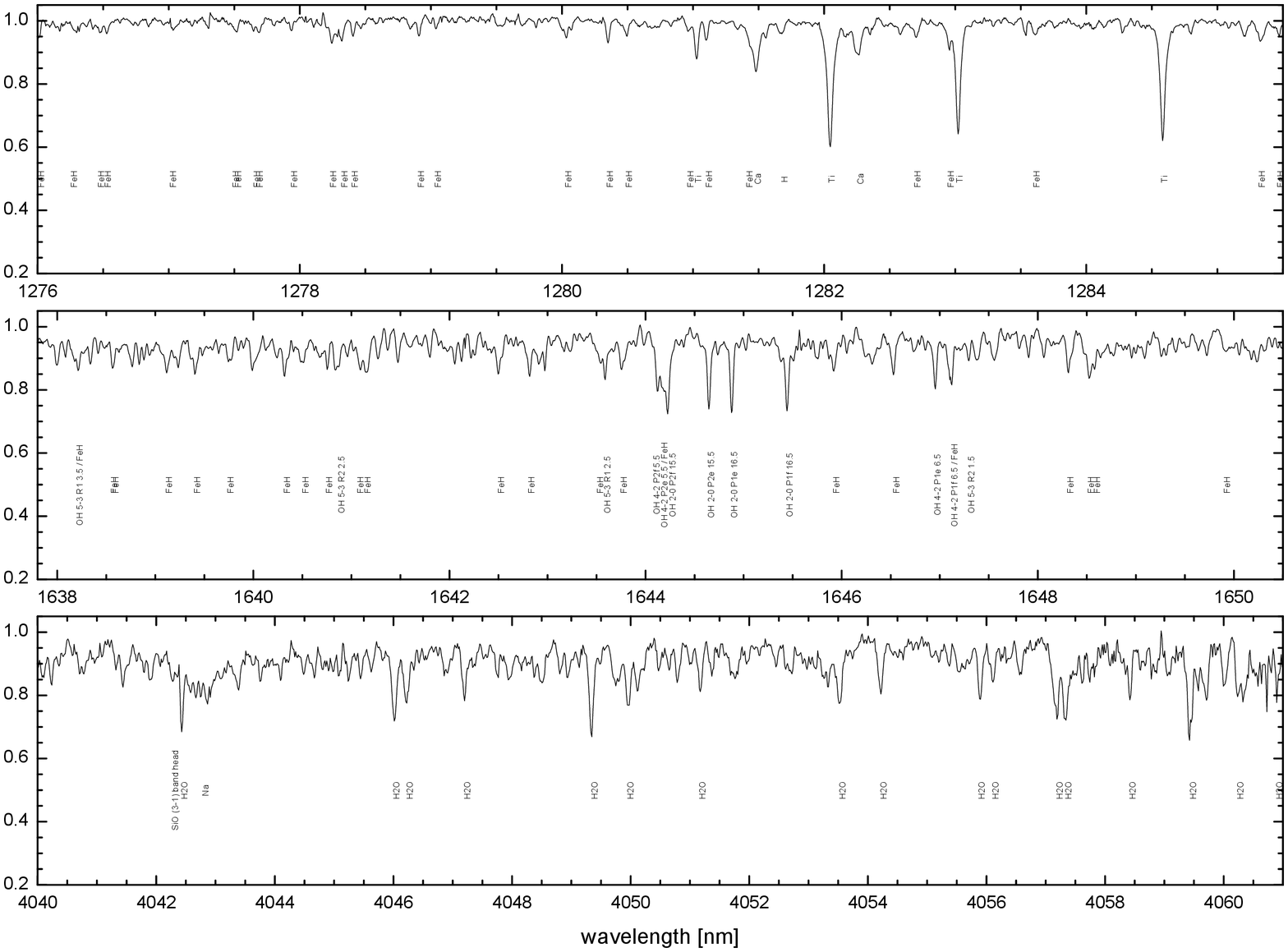}
\centering   
   \caption{Same as Fig.\,\ref{tausco} for Barnard's star (M4 V).
   The FeH lines in the 1280 nm region are from the $F\,^{4}\Pi-X\,^{4}\Pi$
   transition, and in the 1644 nm region from the $E\,^{4}\Pi-A\,^{4}\Pi$ region.}
\label{barnard}
\end{figure}
\end{landscape}

The FeH molecule is a strong opacity source in the red and NIR for
low mass stars, and those lines were readily detectable in our $J$ and $H$ band
spectra of GJ\,699. Lines were identified based on line lists published in
Dulick et al.~(\cite{dulick03}) and Hargreaves et al. (\cite{harg10}), 
where only the strongest lines are marked in the plots shown in Fig.\,\ref{barnard}. FeH is
magnetically sensitive and suspected to weaken with lower surface gravity.
The Zeeman broadening and the low-gravity environment thus greatly
suppress those lines in the sunspot spectra, where they are mostly absent
in the spectral region analysed here. To the contrary, low surface
gravity enhances the line depth of CO, and we found CO second overtone
lines in the sunspot spectra around 1650\,nm, which were absent in GJ\,699.

\begin{landscape}
\begin{figure}
\centering
   \includegraphics[width=1.52\textwidth]{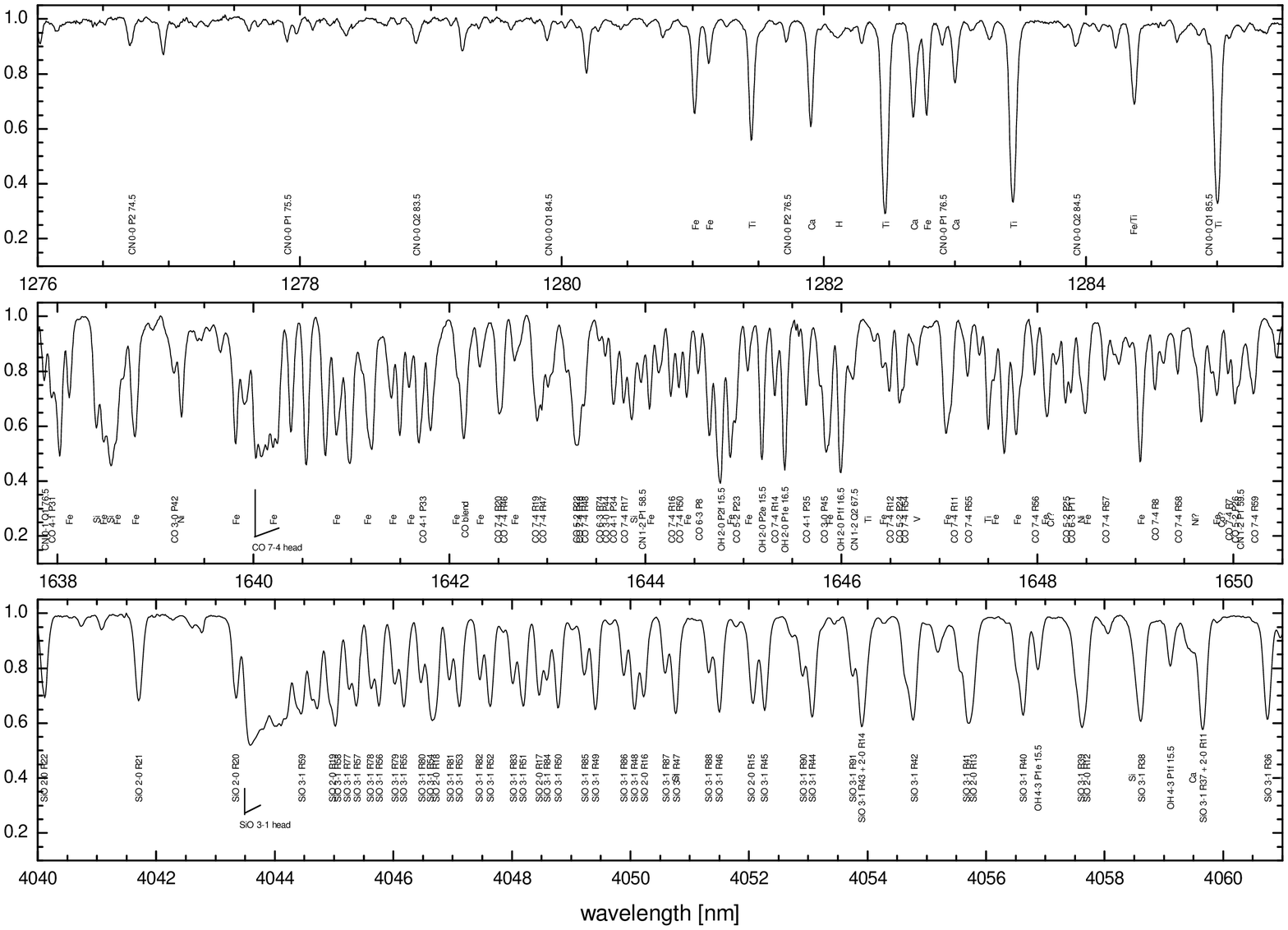}
\centering   
   \caption{Same as Fig.\,\ref{tausco} for HD 61913 (S).}
\label{nzgem}
\end{figure}
\end{landscape}

The 4050\,nm spectrum of GJ\,699 is dominated by a forest of lines, which we
identified as water vapor. In contrast to FeH, where we identified all lines
directly with their counterpart in the line lists, we could only provide a
tentative identification of the H$_{2}$O lines because the latest line lists of
Partridge \& Schwenke (\cite{PS00}) and Barber et al. (\cite{barber06})
provided only a rough match. A number of lines in our
spectrum seem to be limited by the predicted line positions for the same
transition in both lists and a number of predicted lines appear to be missing,
while other lines found in the GJ\,699 spectrum have no unambiguous counterpart in those
lists. All H$_{2}$O lines found in GJ\,699 are present in the FTS sunspot
spectra (although much weaker than in GJ\,699), whereas they remained mostly
unidentified in the Wallace et al. atlas, except for a few lines that are
indeed identified with H$_{2}$O. We note that the main opacity source in the
sunspot spectra around 4050\,nm are SiO lines (some isolated 2-0 lines and
the 3-1 bandhead, see, e.g., the spectrum of NZ Gem, Fig.~\ref{nzgem}), that are, however,
extremely weak in GJ\,699. The same is true for a small number of OH (4-1)
lines falling in this window. The Br\,$\alpha$ line, present in the sunspot
spectra, was undetected in GJ\,699.

\subsubsection{HD 134453 (X TrA)}
The second star where we could not rely on identifications of even
the stronger lines in the literature was the carbon star HD~134453. High resolution
NIR studies on C-stars have lead to interesting results on, e.g.,
the abundance of fluorine (Abia et al. \cite{abia10}, 
Jorissen et al. \cite{joris92}), but a systematic investigation of the
spectral content is still missing. 

In the $J$ and $H$ bands, lines of CN and C$_{2}$ dominate the spectrum. Blending is
a severe problem, hence isolated lines are rare. To determine the contribution of
the various molecules to an individual spectral feature, we calculated synthetic spectra
using the COMARCS models of Aringer et al. (\cite{aringer09}). The line lists of
CN and C$_{2}$ that we used were those from J{\o}rgensen (\cite{jorgen97}) and Querci et al. 
(\cite{QQT74}), respectively. For the model parameters we applied the values of Kipper
(\cite{Kipper04}). Unfortunately, the molecular data in the line lists lack sufficient 
precision for this high resolution study, hence 
the synthetic spectrum differs strongly from the observations 
(see also the discussion in Lebzelter et al. \cite{LZ08} or Lederer et al. \cite{lederer09}). 

\begin{landscape}
\begin{figure}
\centering
   \includegraphics[width=1.52\textwidth]{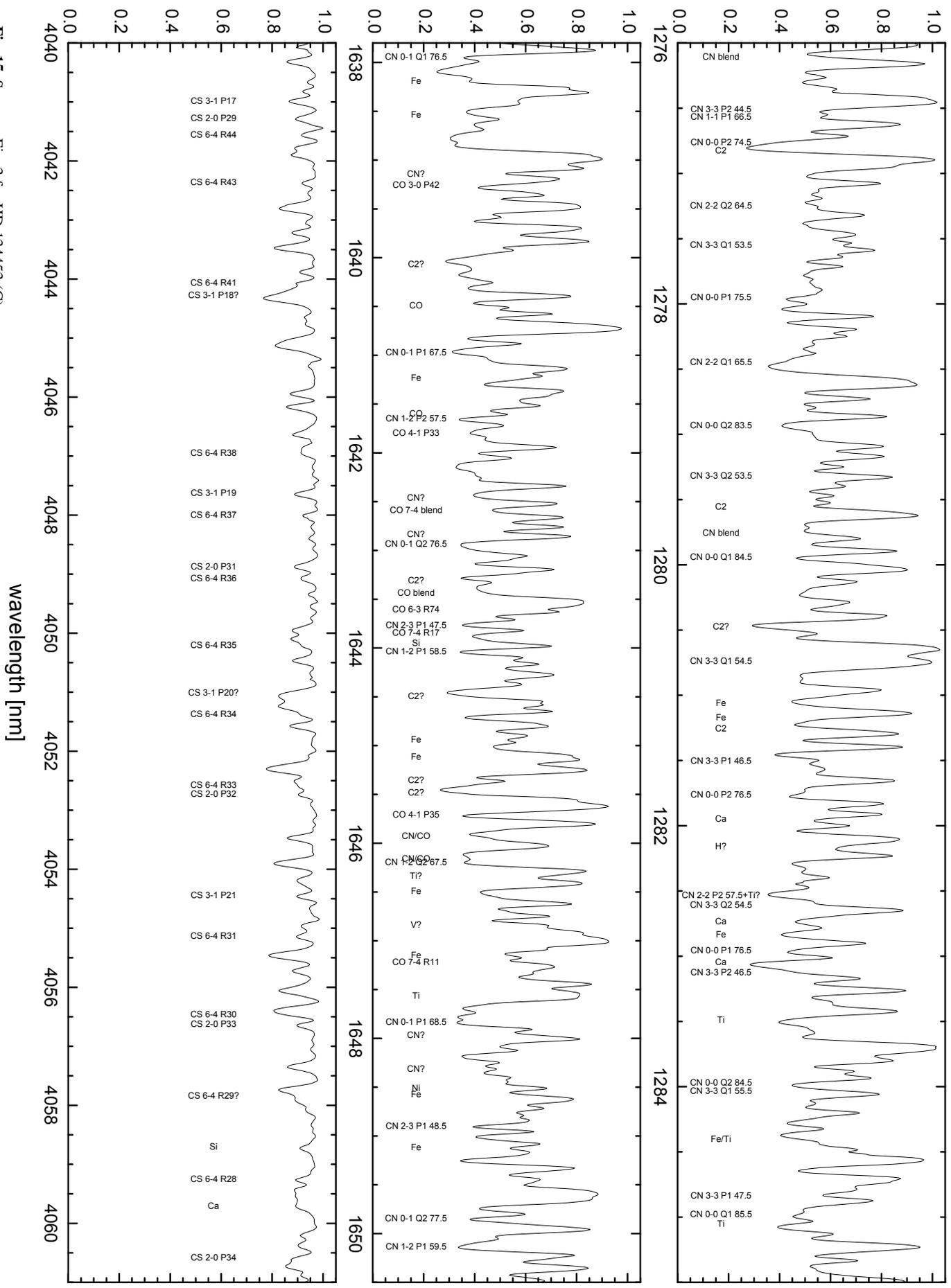}
\centering   
   \caption{Same as Fig.\,\ref{tausco} for HD 134453 (C). }
\label{xtra}
\end{figure}
\end{landscape}

\begin{landscape}
\begin{figure}
\centering
   \includegraphics[width=1.52\textwidth]{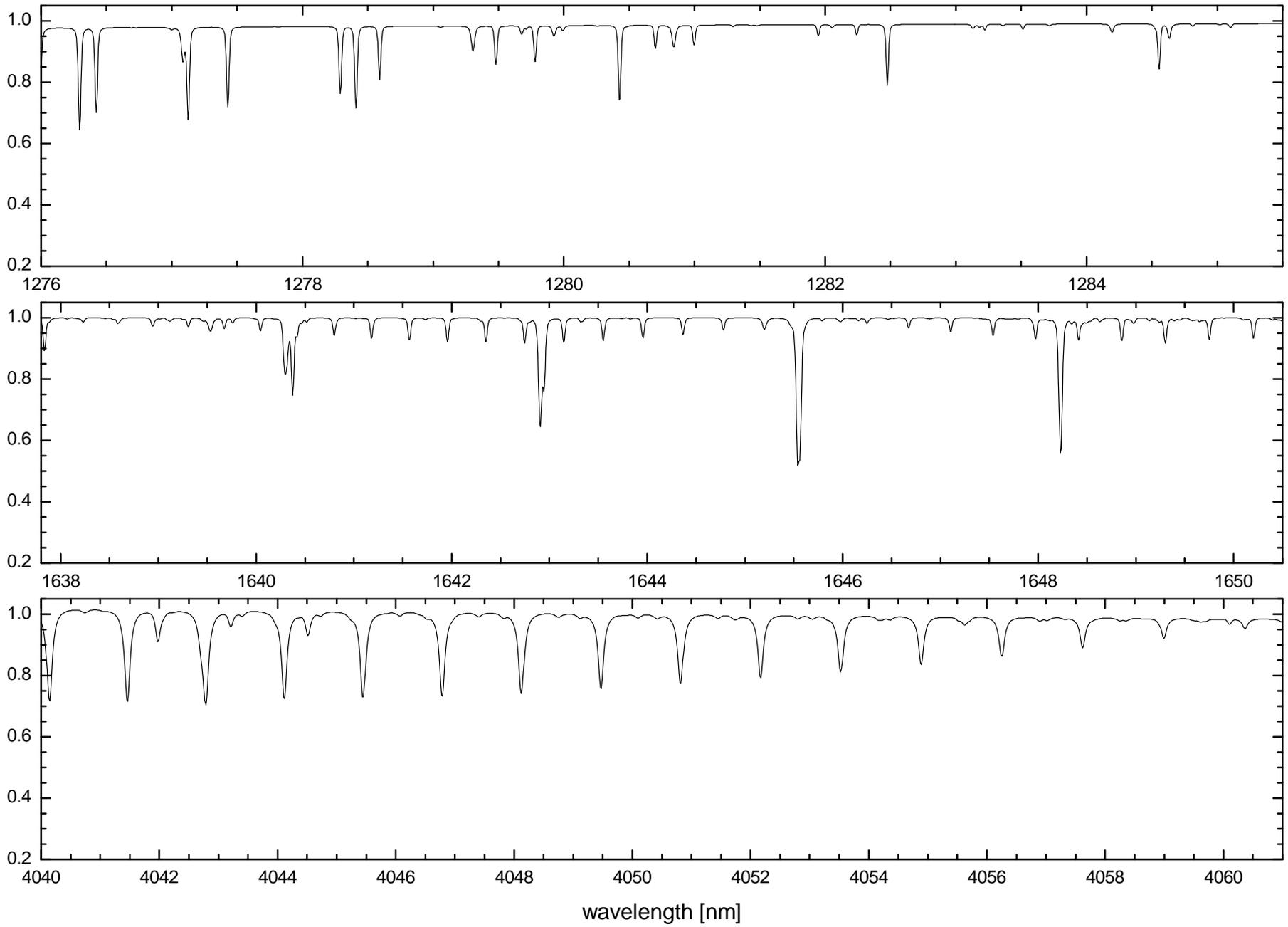}
\centering   
   \caption{Example of a synthetic telluric spectrum used for the correction of CRIRES-POP data.}
\label{telmod}
\end{figure}
\end{landscape}

Focusing on the observed features, which were
at least roughly reproduced by the synthetic spectrum, we re-calculated the model spectra
by removing one molecule (CN, C$_2$, CO) at a time. In
this way, we were able to attribute a number of lines or parts of line blends to a specific
molecule. However, a significant fraction of features could not be identified unambiguously.
For the C$_{2}$ lines (Phillips system), 
the line list did not include identifications of the transitions,
thus it is not provided in Fig.\,\ref{xtra}.

The model was similarly unable to reproduce the section of the spectrum around 4\,$\mu$m. 
Wallace \& Hinkle (\cite{WH02}) identified bands of CH and CS in medium-resolution
spectra of carbon stars in this wavelength range. We were able to attribute
series of lines in our spectrum to transitions of the \mbox{2-0}, \mbox{3-1}, and 
\mbox{6-4} bands of
$^{12}$CS. For this analysis, we used the line lists of Winkel et al. (\cite{cspaper})
and Chandra et al. (\cite{chandra95}). Two likely identifications of atomic lines are also given.

\subsection{Impact on atomic and molecular line lists} 

Even at this preliminary stage, the potential of the 
CRIRES-POP library is clearly evident for an accurate testing of 
existing atomic and molecular line lists. Laboratory and theoretical investigations to create atomic data have in the past
focused on the optical and ultraviolet wavelength regions to support
ground-based and space spectroscopy. Nevertheless, the infrared wavelength
region between approximately 1$\mu$m and 5 $\mu$m contains numerous atomic transitions.
Databases lack transition data for many ions at IR wavelengths, giving the false
impression that this region is uninteresting for studies that are compelled to utilise atomic lines.

Prior to the inauguration of the CRIRES and the Phoenix (Hinkle et al. \cite{phoenix}) instrument, the low spectral
resolution of earlier instruments  did not necessitate accurate
wavelengths or transition probabilities for atomic lines owing to the
effects of line blending\footnote{A limited amount of high spectral resolution data
from instruments such as the Kitt Peak National Observatory Fourier transform
spec\-tro\-meter and the solar spectrum obtained from ATMOS (Geller \cite{geller92})
and the ACE satellite (Hase et al. \cite{hase10}) are available and we clearly need to 
produce
accurate atomic and molecular data.}. In the presence of molecular features typical
of cool stars, atomic lines tend to become incorporated into the broader features.
The incomplete treatment of terrestrial atmospheric lines contributes a
similar effect.

The vast majority of data for atomic transitions at IR wavelengths originates
from atomic structure calculations. These studies were incomplete in terms of
the number of transitions and possess accuracies that remain, in many
cases, undetermined. Modern experimental techniques for determining
oscillator strengths from transition probabilities have been based on the measurement
of energy level lifetimes and branching fractions from line intensities (Wahlgren \cite{wahlgren10}, 
\cite{wahlgren11}), but few NIR lines have been analysed for atomic parameters.

Fortunately, many IR transitions originate from energy levels that are studied
experimentally for optical and ultraviolet transitions. Additional lifetime
measurements will no doubt need to be made to enable the strongest
and/or best placed lines in the NIR of various elements and ions to be analysed. The
large amount of ongoing laboratory analyses should therefore be targeted toward recording
emission line spectra for intensity measurements to determine transition
branching fractions. The wavelength region between 0.9 $\mu$m and 2.5 $\mu$m is
of particular importance for atomic data, as this region contains known
transitions of many post iron-group elements. The iron-group elements are 
of interest at all wavelengths to provide the wavelength standards that can be
used to adjust the CRIRES wavelength scale in regions where
few calibration lines exist, and to apply accurate oscillator strengths to
place abundance analyses at UV, optical, and IR wavelengths on a similar scale.
Finally, the spectral library will provide valuable comparison
data for testing line-broadening theories. Few precise data are
available so far, but they are in high demand for quantitative analyses
using not only the hydrogen and helium lines, but also several 
rather strong metal transitions.

\section{Conclusions and outlook} 
We have presented the first results  
and relevant background information for our high-resolution spectral library in
the infrared called CRIRES-POP. The spectral data that we have 
presented illustrate the potential of our library not only for the studies of 
individual stars that are underway within the CRIRES-POP consortium, but also comparative studies spanning a range of 
spectral types. As an immediate highlight, we considered 
the case of Barnard's star. Our high-resolution spectrum has
allowed for the first time a direct identification of photospheric water vapor lines in a cool M dwarf
and demonstrated the incompleteness of present-day H$_{2}$O line-lists. The large wavelength coverage of the CRIRES-POP observations, combined with
its high spectral resolution, ensures that they provide an invaluable resource for prioritising laboratory measurements of both wavelengths and transition probabilities.
The diversity in the properties of stars observed within the CRIRES-POP program will provide spectral lines with a large range of excitation energies and ionisation
stages. We invite and encourage the scientific community to use the reduced data of our library for further studies of
the individual stars, but also for comparison, model testing, and preparation of observations, but 
kindly ask researchers to give
reference to this publication. 

\acknowledgements
TL acknowledges support by the Austrian Science Fund FWF under project numbers P20046-N16, P21988-N16, and
P23006-N16, and SU under project number P22911-N16. 
HH is supported by the Swedish Research Council (VR) through grant 621-2006-3085.
GMW acknowledges support from NASA Grant NNG06GJ29G. 
SU acknowledges support from the Fund for Scientific Research of Flanders (FWO) under grant number G.0470.07.
ASe acknowledges support from the National Science Foundation under grant NSF AST-0708074.
Major thanks go to the telescope operators and support astronomers at ESO Garching and 
Paranal for doing an excellent job in observing the huge number of wavelength settings necessary for this project.

{}

{\it Note added in proof:} After acceptance of the paper we became aware that we did not use the line
list of Barber et al. (2006) for the identification of water vapor lines
in GJ699 but by mistake the line list in HITEMP 1995 (Rothman et al 1995).
HITEMP 1995 is superseded by the greatly improved HITEMP 2010 (Rothman et
al. 2010) database. We regret this mistake and have to revise our
statement about the incompleteness of present day water vapor line lists
which is thus unfounded. Nevertheless, we are convinced that the
CRIRES-POP spectral library will offer an excellent possibility to perform
detailed comparisons between theoretical line lists and experimental data.

%\end{landscape}

\end{document}